%% file: main.tex
\pdfoutput=1
\documentclass[11pt]{article}
\input{preamble}

\title{\textbf{Benchmarking Classification Tasks of Chest X-ray AI Across
Multinational Public and Private Datasets: a retrospective diagnostic accuracy study}}

\author[1,2,$\dagger$]{Qinmei Xu}
\author[1,$\dagger$]{Yiheng Li}
\author[3]{Wenya Zhao}
\author[4]{Zihao Zhao}
\author[5]{Chuanjun Xu}
\author[1]{Ahmet Gorkem Er}
\author[1]{Xianghao Zhan}
\author[6]{Brittany Dashevsky}
\author[7]{Mohammed Alawad}
\author[4]{Marc von der St\"uck}
\author[4]{Marvin Gazibari\'c}
\author[4]{Daniel Truhn}
\author[3]{Shunsuke Shibata}
\author[3]{Ryosuke Horino}
\author[3]{Mio Ito}
\author[3]{Akio Hiwatashi}
\author[8]{Suyu Li}
\author[8]{Mengya Yang}
\author[8]{Ya Liu}
\author[8]{Zhiyuan Sun}
\author[8]{Changsheng Zhou}
\author[2,*]{Haruka Itakura}
\author[8,*]{Xiao Li}
\author[1,*]{Olivier Gevaert}

\affil[1]{Stanford Division of Computational Medicine, Department of Medicine, Stanford University, Stanford, CA, USA}
\affil[2]{Division of Oncology, Department of Medicine, Stanford University, Stanford, CA, USA}
\affil[3]{Department of Radiology, Nagoya City University Graduate School of Medical Sciences, Nagoya, Japan}
\affil[4]{Department of Diagnostic and Interventional Radiology, University Hospital Aachen, Aachen, Germany}
\affil[5]{Department of Radiology, the Second Hospital of Nanjing, Nanjing University of Chinese Medicine, Nanjing, China}
\affil[6]{Stanford Breast Imaging, Department of Radiology, Stanford University, Stanford, CA, USA}
\affil[7]{National Center for AI (NCAI), Saudi Data and AI Authority (SDAIA), Riyadh, Saudi Arabia}
\affil[8]{Department of Radiology, Jinling Hospital, Nanjing, Jiangsu, China}
\affil[$\dagger$]{These authors contributed equally to this work.}
\affil[*]{Corresponding authors: Haruka Itakura (\href{mailto:itakura@stanford.edu}{itakura@stanford.edu}), Xiao Li (\href{mailto:15050528100@yeah.net}{15050528100@yeah.net}), Olivier Gevaert (\href{mailto:ogevaert@stanford.edu}{ogevaert@stanford.edu}).}

\date{}

\begin{document}
\maketitle

\input{sections/00_summary}

\clearpage
\input{sections/00b_research_in_context}

\clearpage
\input{sections/01_introduction}
\input{sections/02_methods}

\input{sections/03_results}
\input{sections/04_discussion}

\input{sections/06_statements}

\clearpage
\bibliographystyle{vancouver}
\bibliography{refs}

\clearpage
\setcounter{section}{0}
\setcounter{figure}{0}
\setcounter{table}{0}
\renewcommand{\thesection}{S\arabic{section}}
\renewcommand{\thesubsection}{S\arabic{section}.\arabic{subsection}}
\renewcommand{\thefigure}{S\arabic{figure}}
\renewcommand{\thetable}{S\arabic{table}}
\renewcommand{\theHsection}{S\arabic{section}}
\renewcommand{\theHsubsection}{S\arabic{section}.\arabic{subsection}}
\renewcommand{\theHfigure}{S\arabic{figure}}
\renewcommand{\theHtable}{S\arabic{table}}
\crefname{section}{appendix section}{appendix sections}
\Crefname{section}{Appendix section}{Appendix sections}

\begin{center}
  {\Large\bfseries Appendix}
\end{center}
\vspace{0.5em}

\noindent
This appendix contains appendix sections S1--S8, tables S1--S5, and figure S1.
References to ``figure'' and ``table'' without an S prefix refer to the main Article.

\input{sections/S0_cohort_selection}
\input{sections/S0b_reference_standard}
\input{sections/S1_harmonization}

\input{sections/S2_models}
\input{sections/S3_autogluon}
\input{sections/S4_performance}
\input{sections/S4b_spatial_extent}
\input{sections/S5_prevalence}

\end{document}

%% file: preamble.tex
\usepackage[utf8]{inputenc}
\usepackage[T1]{fontenc}
\usepackage[margin=1in]{geometry}
\usepackage{graphicx}
\usepackage{adjustbox}
\usepackage{booktabs}
\usepackage{array}
\usepackage{multirow}
\usepackage{longtable}
\usepackage{tabularx}
\usepackage{pdflscape}
\usepackage{ragged2e}
\usepackage{siunitx}
\usepackage{amsmath,amssymb}
\usepackage{authblk}
\usepackage[table,dvipsnames]{xcolor}
\usepackage[super,sort&compress,numbers]{natbib}
\usepackage{microtype}
\usepackage{caption}
\usepackage[most]{tcolorbox}
\usepackage{hyperref}
\usepackage[capitalize]{cleveref}   %

\definecolor{rowband}{gray}{0.92}
\definecolor{pubcol}{HTML}{0072B2}
\definecolor{pricol}{HTML}{D55E00}
\definecolor{prevcol}{HTML}{2C7FB8}

\newcolumntype{L}[1]{>{\RaggedRight\arraybackslash}p{#1}}

\graphicspath{{figures/}}

\crefname{figure}{figure}{figures}
\Crefname{figure}{Figure}{Figures}
\crefname{table}{table}{tables}
\Crefname{table}{Table}{Tables}

\newtcolorbox{researchincontext}{
  colback=white, colframe=black, boxrule=0.8pt, arc=0pt,
  left=10pt, right=10pt, top=10pt, bottom=10pt,
  breakable, enhanced jigsaw,
  title={\bfseries\large Research in context}, coltitle=black, colbacktitle=white,
  fonttitle=\bfseries
}
\newcommand{\ricsection}[1]{\par\medskip\noindent\textbf{#1}\par\nobreak\smallskip}

\newif\ifshowtodo\showtodotrue

%% file: sections/00_summary.tex
\begin{abstract}
\noindent\textbf{Background}\quad
Chest X-ray (CXR) artificial intelligence models are increasingly proposed for triage, screening, and rule-out workflows, in which the operating point is fixed so that few abnormalities are missed. Most published evaluations instead report threshold-free discrimination on a few public benchmarks, leaving clinically usable performance and generalisability to unseen institutional cohorts poorly characterised.

\medskip\noindent\textbf{Methods}\quad
In this retrospective multinational study we evaluated 10 published CXR classification models on 12 datasets from six countries: seven public benchmarks and five institutional cohorts from China, Japan, and Germany. Models spanned supervised convolutional networks, vision-language models, a question-answering foundation model, and a repurposed encoder. Only image-level classification was assessed; segmentation, lesion detection, and report generation were out of scope. Source labels were harmonised into a common finding vocabulary. The primary outcome was specificity at 95\% sensitivity; positive predictive value (PPV) at 95\% sensitivity and AUROC were secondary outcomes. Estimates were aggregated with equal weight per diagnostic task, with 95\% CIs.

\medskip\noindent\textbf{Findings}\quad
MAVL and MedKLIP achieved the highest overall specificity at 95\% sensitivity (0\textperiodcentered{}37 [95\% CI 0\textperiodcentered{}31--0\textperiodcentered{}44] and 0\textperiodcentered{}37 [0\textperiodcentered{}30--0\textperiodcentered{}44]) and kept that ranking when restricted to tasks evaluable for every model. PPV at the same operating point was 0\textperiodcentered{}18 (0\textperiodcentered{}13--0\textperiodcentered{}24) and 0\textperiodcentered{}19 (0\textperiodcentered{}13--0\textperiodcentered{}25), so most flagged examinations would be false positives. Specificity was lower on institutional cohorts than on public benchmarks for most models, by 0\textperiodcentered{}13 for MAVL, 0\textperiodcentered{}15 for DenseNet, and 0\textperiodcentered{}09 for MedKLIP. No model was best for every finding, country, lesion extent, or prevalence stratum; specificity fell from 0\textperiodcentered{}34 for diffuse to 0\textperiodcentered{}15 for focal findings.

\medskip\noindent\textbf{Interpretation}\quad
Calibrated to miss few abnormalities, the best current CXR classification models clear only about a third of truly negative examinations, and public-benchmark performance overstates what the same models achieve on unseen institutional data. Operating-point-aware evaluation and external validation on institutional cohorts should precede deployment.

\medskip\noindent\textbf{Funding}\quad
National Natural Science Foundation of China.
\end{abstract}

%% file: sections/00b_research_in_context.tex
\begin{researchincontext}

\ricsection{Evidence before this study}
We searched PubMed/MEDLINE, Europe PMC, and arXiv from database inception to Aug 27, 2026, using combinations of ``chest radiograph'', ``chest x-ray'', ``artificial intelligence'', ``deep learning'', ``foundation model'', ``vision-language model'', ``benchmark'', ``external validation'', ``triage'', ``rule-out'', ``operating point'', and ``long-tailed''. The PubMed searches were restricted to English-language publications. We screened 1295 records.

Previous studies developed or externally validated supervised multi-label systems and vision-language foundation models, but generally assessed a single model or a small number of datasets; only two retrieved studies spanned both public benchmarks and private institutional cohorts, and each validated a single system, though multicohort validation showed performance varying across institutions, acquisition settings, patient populations, and reference standards. Head-to-head studies compared commercial algorithms for tuberculosis or a small number of common radiographic findings and showed threshold-dependent performance and variable false-positive burdens; of the 10 retrieved, none compared more than seven systems outside a single country or covered more than five findings. Triage and autonomous-reporting studies estimated workload reduction at very high sensitivity, but generally assessed a single system within a narrowly defined workflow. Long-tailed and zero-shot evaluations broadened disease coverage, but relied predominantly on public benchmarks and threshold-independent ranking metrics. Only one retrieved study reported specificity at 95\% sensitivity for more than one system, and none reported positive predictive value at that operating point.

The retrieved evidence is at substantial risk of bias: nine of the 10 head-to-head comparisons were retrospective, sampling was frequently enriched or case-control rather than consecutive, and systematic reviews report high risk of bias in most included studies---66\% of 46 chest radiograph AI lung-cancer studies in at least one domain---with heterogeneous and often inadequately reported reference standards. The systems and target conditions are too heterogeneous for a meaningful pooled estimate, and none was attempted. We did not identify a peer-reviewed study combining multiple contemporary model paradigms, geographically diverse public and institutional cohorts, harmonised finding-level labels, a common high-sensitivity target, and matched shared-task analyses.

\ricsection{Added value of this study}
This retrospective multinational benchmark compared ten chest x-ray artificial intelligence model configurations spanning supervised convolutional, vision-language, question-answering, and repurposed foundation-model approaches across 12 datasets from six countries on three continents, comprising 466\,135 evaluated radiographs---nearly 20 times the largest previous head-to-head comparison (23\,954 people)---and 27 diagnostic tasks annotated in two or more cohorts. The benchmark included 385\,559 radiographs from model-unseen institutional cohorts, including one cohort that became publicly available only after the original evaluation and four restricted cohorts evaluated on site.

Nine models producing continuous scores were evaluated at operating points corresponding to 95\% sensitivity, and the categorical question-answering model was assessed separately, with coverage-aware and matched shared-task analyses addressing unequal label coverage across models and datasets. Knowledge-enhanced vision-language models ranked highest overall, but the best equal-weight task-averaged specificity at 95\% sensitivity was approximately 0\textperiodcentered{}37, with a positive predictive value of approximately 0\textperiodcentered{}19, so about four in five flagged examinations would be false positives. Several leading models performed worse on model-unseen institutional cohorts than on established public benchmarks, with specificity lower by 0\textperiodcentered{}13 for MAVL, 0\textperiodcentered{}15 for DenseNet, and 0\textperiodcentered{}09 for MedKLIP; no model was uniformly best across datasets or radiographic findings; and focal findings were more difficult than diffuse findings for most models (0\textperiodcentered{}15 vs 0\textperiodcentered{}34).

Because the released evaluation framework takes a new model through a wrapper around its inference call, a system submitted by its developers can be scored on the same 12 cohorts---public benchmarks and restricted institutional cohorts alike---with no patient-level radiographs or labels leaving any institution, allowing the benchmark to be extended to models published after this Article.

\ricsection{Implications of all the available evidence}
High discrimination on familiar public benchmarks, or high sensitivity in a single use case, is insufficient to establish deployment readiness. Chest x-ray artificial intelligence should be evaluated on data independent of model development, at prespecified operating points appropriate for the intended clinical task, with transparent cohort-specific reference standards and task-matched comparisons.

These finding-level results do not establish examination-level autonomous reporting or support treating current systems as interchangeable standalone rule-out tools. Prospective studies should assess fixed-threshold transfer, local calibration, subgroup performance, examination-level endpoints, workload, diagnostic errors, effects on patient management, and safety.

\end{researchincontext}

%% file: sections/01_introduction.tex
\section{Introduction}\label{sec:intro}

Chest radiography remains central to the evaluation of cardiopulmonary disease, trauma, device placement, and infectious processes. Because CXRs are among the most frequently performed imaging examinations worldwide, increasing imaging volume and radiology workforce constraints have motivated the development of automated interpretation systems for triage, screening, and clinical decision support.\citep{hosny2018airadiology} Early deep-learning approaches largely relied on supervised convolutional neural networks trained to predict a fixed set of radiographic labels.\citep{rajpurkar2017chexnet,wang2017chestxray8} More recent vision-language foundation models use paired radiographs and reports, textual disease descriptions, prompt learning, or knowledge-enhanced alignment to support broader and more flexible disease recognition.\citep{tiu2022chexzero,bannur2023biovilt,wu2023medklip}

Despite rapid progress, the clinical utility and generalizability of CXR AI models remain incompletely understood. Many prior studies have relied primarily on global discrimination metrics such as AUROC and have evaluated models on a limited number of public benchmarks.\citep{rajpurkar2018chexnext} However, deployment does not occur across all possible thresholds; it occurs at a specific operating point selected for a clinical use case. For CXR AI systems intended for screening, triage, or rule-out workflows, minimizing missed abnormalities is typically the first priority. We therefore selected specificity at 95\% sensitivity as a pragmatic and clinically interpretable high-sensitivity benchmark for comparative evaluation. Prior CXR AI studies have prespecified 95\% sensitivity for general abnormality detection and disease-specific settings,\citep{Nabulsi2021NormalAbnormal} or evaluated sensitivity thresholds of approximately 95\% and higher for rule-out and automatic-reporting workflows.\citep{Plesner2024Radiology,Schalekamp2024EurRadiol} The optimal sensitivity threshold nevertheless remains task- and context-dependent and may need to be higher when the clinical consequences of missed findings are substantial.

Operating-point performance, however, is only informative if it holds beyond the data on which a model was built. A second major limitation of existing evaluations is therefore the lack of systematic external validation. Most current models are developed and evaluated primarily on the same public datasets and may not generalize to new private clinical cohorts because of differences in patient populations, image acquisition protocols, disease prevalence, labeling definitions, and institutional practice patterns.\citep{zech2018confounding} Similarly, performance may differ across diagnostic categories. Focal abnormalities such as nodules, fractures, or small pneumothoraces may be more difficult to detect than diffuse findings such as edema or widespread opacities. Rare findings may also expose weaknesses that are not apparent when performance is averaged across common abnormalities.

To address these gaps, we performed a large-scale multinational benchmark of 10 CXR classification models on public and private datasets from six countries. All 10 systems were benchmarked on a single task type: image-level classification, in which the model returns a score for the presence of each radiographic finding. Segmentation, lesion detection, and report generation were outside the scope of this study, so the conclusions drawn here concern CXR classification models and not CXR AI across all task types. We evaluated models using clinically oriented operating-point metrics, with emphasis on specificity at 95\% sensitivity, and characterized heterogeneity by public versus private data source, country, diagnostic task, spatial disease extent, and disease prevalence. Our goal was to determine not only which models perform best overall, but also whether their performance is clinically usable and where they fail.

%% file: sections/02_methods.tex
\section{Methods}\label{sec:methods}

\begin{figure}[tp]
    \centering
    \includegraphics[width=\linewidth]{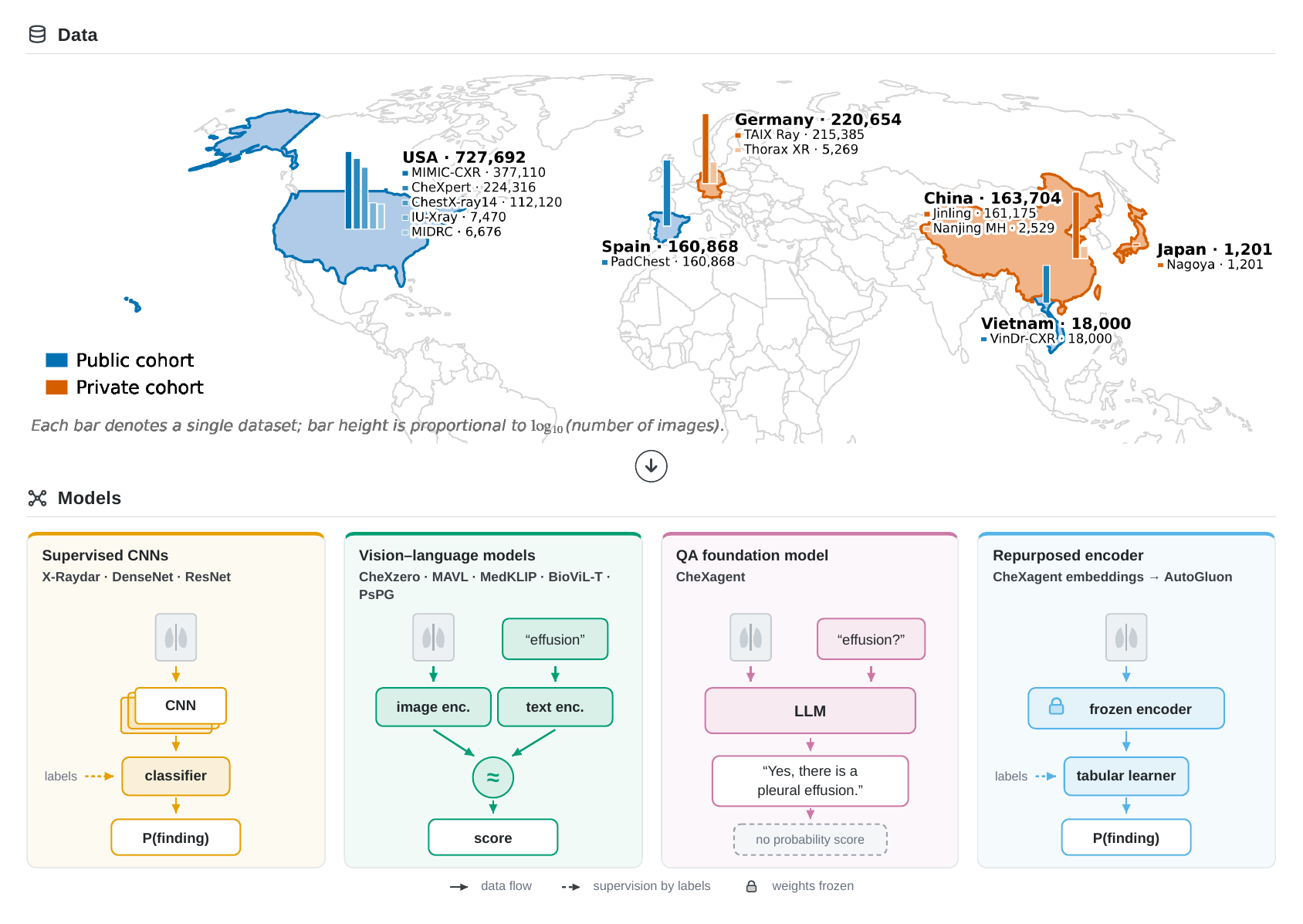}
    \caption{\textbf{Study design: cohorts and model paradigms.} \textbf{Top}, geographic distribution of the 12 chest-radiograph cohorts; each bar denotes one dataset, bar height is proportional to $\log_{10}$(number of images), and colour denotes public (blue) versus private (orange) access. \textbf{Bottom}, the four model paradigms evaluated: supervised CNN classifiers trained directly on labelled radiographs; vision-language models scored zero-shot by image--text embedding similarity; a QA-based foundation model that emits a free-text answer and therefore no continuous score; and a repurposed encoder in which frozen CheXagent embeddings are passed to a supervised tabular learner. Solid arrows denote data flow, dashed arrows denote supervision by labels, and the lock denotes frozen weights.}
    \label{fig:overview}
\end{figure}

\subsection{Ethical approval}

This retrospective study was conducted in accordance with the Declaration of Helsinki.

Use of the Nanjing Multi-Hospital cohort was approved by the Ethics Committee of the Second Hospital of Nanjing, Nanjing University of Chinese Medicine (approval no.~2026-LS-ky008); this single approval covers all participating hospitals and all subsets of the cohort. Informed consent for the research use of the imaging data was obtained from all participants or, for participants younger than 18 years, from a parent or legal guardian.

Use of the Nagoya cohort was approved by the Medical Research Ethics Review Committee of Nagoya City University Graduate School of Medical Sciences (approval no.~60-25-0068; approved 3 August 2025).

For the German cohorts (Thorax XR and TAIX Ray), the Ethics Committee of the Medical Faculty of RWTH Aachen University determined that formal ethics approval was not required for this retrospective study and raised no ethical or professional concerns (reference no.~EK 205/17).

For the Jinling cohort, the radiographs were retrieved retrospectively from the clinical imaging archive of Jinling Hospital and did not leave the institution; model inference was performed on site, and only anonymized per-image model outputs and reference labels were transferred to the analysing institutions. Because the study was conducted exclusively on anonymized data, it was exempt from ethics committee review under the applicable national regulations, and the requirement for informed consent was waived.

All radiographs were de-identified before analysis, and no protected health information was transferred between institutions. The seven public benchmark datasets were used under their respective data use agreements and required no additional ethical approval.

\subsection{Study design, datasets, and label harmonization}

All 10 systems were evaluated on a single task type: image-level classification of radiographic findings, in which a model returns a score for each harmonized finding and that score is compared with the cohort's reference label for the same finding. No model was scored on segmentation masks, lesion bounding boxes, or generated reports; CheXagent, the only evaluated model that natively emits free text, was scored by parsing its answer into a yes/no label. Throughout this paper, \emph{task} denotes one diagnostic label evaluated within one dataset, and not a task type in the machine-learning sense.

We conducted a retrospective multinational benchmarking study of chest X-ray (CXR) AI models using both public benchmark datasets and private clinical cohorts; the overall study design and analysis workflow are summarized in \cref{fig:overview}. The benchmark included 12 datasets from six countries. Public datasets included CheXpert,\citep{irvin2019chexpert} MIMIC-CXR,\citep{johnson2019mimic} ChestX-ray14 (Google),\citep{wang2017chestxray8} MIDRC,\citep{tsai2021ricord} IU-Xray,\citep{demnerfushman2016iuxray} PadChest,\citep{bustos2020padchest} and VinDr-CXR.\citep{nguyen2022vindrcxr} Private cohorts included Jinling, Nanjing Multi-Hospital, Nagoya, Thorax XR, and TAIX Ray. The TAIX Ray cohort has since become publicly available: it was deposited in a public repository in February 2026 under a CC~BY~4.0 licence and described in a data descriptor published in April 2026.\citep{truhn2026taixray} That release post-dates the development of all 10 evaluated models, the most recent of which was published in 2024, so no evaluated model can have been trained on these radiographs and the cohort was not available as a benchmark while those models were being built. Within this benchmark TAIX Ray therefore behaves as the other institutional cohorts do---an external cohort unseen by every model tested---and we retain it in the private group on that basis; the grouping reflects the cohort's role in this evaluation rather than its current distribution status. The four remaining private cohorts (Jinling, Nanjing Multi-Hospital, Nagoya, and Thorax XR) are not publicly available. Only frontal chest radiographs were included when view information was available. For the five private cohorts, reference labels were ascertained locally and finalized before benchmark model inference, either by radiologist assessment of the radiograph together with the finalized clinical report (Nagoya, Nanjing Multi-Hospital) or from structured diagnostic fields of routine clinical reports (Jinling, TAIX Ray, Thorax XR); no model output was used to create, revise, or adjudicate any reference label, and uncertain or indeterminate source values were coded as missing rather than negative. The cohort-specific ascertainment procedures are described in the appendix. Imaging labels were harmonized across pulmonary, pleural, cardiac, mediastinal, skeletal, diaphragmatic, and device-related findings. Synonymous or overlapping labels were merged when clinically appropriate to support cross-dataset comparison. An overview of the datasets, their provenance, and label coverage is given in \cref{tab:dataset-overview}, which lists the 27 diagnostic tasks annotated by at least two cohorts; the complete grid, including the 25 tasks annotated by a single cohort, is given in the appendix.

\input{tables/table1_dataset_overview}

\subsection{Evaluated models}

We evaluated 10 CXR diagnostic models representing four major learning paradigms. Supervised CNN classifiers included X-Raydar,\citep{cid2024xraydar} ResNet,\citep{he2016resnet} and DenseNet;\citep{huang2017densenet} the ResNet and DenseNet checkpoints were obtained from the TorchXRayVision library.\citep{cohen2022torchxrayvision} Vision-language foundation models included CheXzero,\citep{tiu2022chexzero} BioViL-T,\citep{bannur2023biovilt} MAVL,\citep{phan2024mavl} MedKLIP,\citep{wu2023medklip} and PsPG.\citep{ye2024pspg} CheXagent\citep{chen2024chexagent} was evaluated as a QA-based foundation model. CheXagent-AutoGluon was evaluated as a repurposed encoder model in which CheXagent image embeddings were used for downstream supervised classification with probability output via AutoGluon,\citep{erickson2020autogluon} an approach consistent with prior work that repurposes learned lesion representations for downstream prediction.\citep{li2024lungvae} The evaluated models, their paradigms, architectures, the data each was trained on, their training objectives, inference interfaces, and source references are summarized in \cref{tab:models}. Each model was evaluated at its published release, with its original weights and without fine-tuning; the checkpoints, the text prompts used by the zero-shot vision-language models and by CheXagent, and the CheXagent-AutoGluon training configuration are contained in the analysis code released with this Article (see Data sharing). No generative artificial intelligence tool was used to produce, select, or adjudicate any result reported here.

\input{tables/table2_models}

\subsection{Outcomes and statistical analysis}

The primary clinically oriented metric was specificity at 95\% sensitivity. This metric was selected to reflect high-sensitivity screening, triage, and rule-out workflows, in which minimizing missed abnormalities is prioritized. At a fixed 95\% sensitivity threshold, specificity quantifies the proportion of truly negative examinations that can still be correctly excluded. Thus, higher specificity at this threshold indicates greater potential to reduce radiologist workload while maintaining a low miss rate.

The unit of analysis was a dataset-label-model unit. For each model, performance was estimated for each available label within each dataset. Dataset-label units lacking at least one positive and one negative example, for which an operating point is undefined, were excluded from operating-point estimation. AUROC and PPV at 95\% sensitivity were evaluated as complementary metrics. Overall model performance was aggregated using equal-weight per task to prevent tasks represented in more datasets from dominating the summary. Each dataset-label unit's metrics were accompanied by 95\% confidence intervals from 1{,}000 bootstrap resamples of the evaluation set. Confidence intervals for the equal-weight aggregated summaries were obtained from a normal approximation across tasks, as the mean of the per-task values $\pm$\,1\textperiodcentered{}96 standard errors, where the standard error is the standard deviation of the per-task means divided by the square root of the number of contributing tasks. Because these summaries average over a few dozen tasks, this interval is closely approximated by resampling tasks with replacement: across the reported summaries the two approaches agree to within 0\textperiodcentered{}005.

Because models differed in task coverage and output availability, we performed both a coverage-aware overall analysis across all available dataset-task units and a fair shared-task analysis restricted to dataset-task units available for all probabilistic models (all evaluated models except the QA-based CheXagent, which emits categorical yes/no answers rather than calibrated probability scores). Two complementary summaries were therefore reported. The model-ranking figures show the full distribution of per-unit values, with models ordered by the median across units, so that between-unit variability remains visible. All summary point estimates and stratified analyses instead use the equal-weight-per-task aggregation described above, in which values are averaged across datasets within a task and then across tasks. Additional stratified analyses assessed public versus private data source, country, diagnostic task, spatial extent of the target finding, and disease prevalence. The public-private generalization gap was calculated as private-cohort performance minus public-benchmark performance.

\subsection{Role of the funding source}

The funder of the study had no role in study design, data collection, data analysis, data interpretation, or writing of the report. QX and YiL had full access to all the de-identified derived data in the study; access to the underlying radiographs of each institutional cohort was held by that cohort's site investigators, who performed model inference locally and transferred only de-identified per-image model outputs and reference labels. The corresponding authors had final responsibility for the decision to submit for publication.

%% file: tables/table1_dataset_overview.tex
\begin{table}[tbp]
\centering
\caption{Overview of the chest-radiograph datasets in the benchmark: access, provenance, image format, evaluated cohort size, and the per-task prevalence (\% positive in the evaluated cohort) of each imaging label. Restricted to the 27 diagnostic tasks evaluated in at least two cohorts; the appendix gives the complete grid, including the 25 tasks annotated by a single cohort.}
\label{tab:dataset-overview}
\scriptsize
\setlength{\tabcolsep}{3.5pt}
\renewcommand{\arraystretch}{1.0}
\begin{adjustbox}{max width=\textwidth,max totalheight=0.82\textheight,center}
\begin{tabular}{>{\raggedright\arraybackslash}p{3.0cm} !{\color{black!18}\vrule} c c c c c c c c c c c c}
\toprule
 & \multicolumn{7}{c}{\textcolor{pubcol}{\textsc{Public}}} & \multicolumn{5}{c}{\textcolor{pricol}{\textsc{Private}}} \\
\cmidrule(lr){2-8}\cmidrule(lr){9-13}
\textbf{Dataset} & \rotatebox{90}{\textbf{CheXpert}} & \rotatebox{90}{\textbf{MIDRC}} & \rotatebox{90}{\textbf{ChestX-ray14 (Google)}} & \rotatebox{90}{\textbf{IU-Xray}} & \rotatebox{90}{\textbf{PadChest}} & \rotatebox{90}{\textbf{VinDr-CXR}} & \rotatebox{90}{\textbf{MIMIC-CXR}} & \rotatebox{90}{\textbf{Nagoya}} & \rotatebox{90}{\textbf{Jinling}} & \rotatebox{90}{\textbf{Nanjing Multi-Hospital}} & \rotatebox{90}{\textbf{TAIX Ray}} & \rotatebox{90}{\textbf{Thorax XR}} \\
\midrule
\textbf{Country} & USA & USA & USA & USA & Spain & Vietnam & USA & Japan & China & China & Germany & Germany \\
\textbf{Image format} & JPEG & DICOM & PNG & PNG & PNG & DICOM & JPEG & DICOM & DICOM & DICOM & PNG & DICOM \\
\textbf{Images, total} & 224{,}316 & 6{,}676 & 112{,}120 & 7{,}470 & 160{,}868 & 18{,}000 & 377{,}110 & 1{,}201 & 161{,}175 & 2{,}529 & 215{,}385 & 5{,}269 \\
\textbf{Images, frontal} & 191{,}010 & 6{,}676 & 112{,}120 & \textemdash{} & \textemdash{} & 18{,}000 & 243{,}334 & 1{,}201 & 161{,}175 & 2{,}529 & 215{,}385 & 5{,}269 \\
\textbf{Official test, $n$} & 668 & \textemdash{} & 1{,}962 & \textemdash{} & \textemdash{} & 3{,}000 & 5{,}159 & \textemdash{} & \textemdash{} & \textemdash{} & \textemdash{} & \textemdash{} \\
\textbf{Evaluated, $n$} & \textbf{403} & \textbf{6{,}676} & \textbf{1{,}695} & \textbf{2{,}561} & \textbf{62{,}838} & \textbf{3{,}000} & \textbf{3{,}403} & \textbf{1{,}201} & \textbf{161{,}175} & \textbf{2{,}529} & \textbf{215{,}385} & \textbf{5{,}269} \\
\midrule
\multicolumn{13}{l}{\itshape Per-task prevalence (\% positive in evaluated cohort; blank = label not provided; ordered by number of datasets; tasks evaluated in $\geq$2 cohorts)} \\
\addlinespace[1pt]
Fracture & \cellcolor{prevcol!11}1\textperiodcentered{}2 &  & \cellcolor{prevcol!19}3\textperiodcentered{}5 & \cellcolor{prevcol!16}2\textperiodcentered{}4 & \cellcolor{prevcol!16}2\textperiodcentered{}7 &  & \cellcolor{prevcol!36}13 & \cellcolor{prevcol!50}25 & \cellcolor{prevcol!20}4\textperiodcentered{}2 & \cellcolor{prevcol!30}8\textperiodcentered{}9 & \cellcolor{prevcol!42}18 & \cellcolor{prevcol!62}82 \\
\rowcolor{rowband}Pleural Effusion & \cellcolor{prevcol!45}20 &  &  & \cellcolor{prevcol!22}4\textperiodcentered{}9 & \cellcolor{prevcol!17}2\textperiodcentered{}9 & \cellcolor{prevcol!19}3\textperiodcentered{}7 & \cellcolor{prevcol!62}55 & \cellcolor{prevcol!60}36 & \cellcolor{prevcol!42}18 & \cellcolor{prevcol!33}11 & \cellcolor{prevcol!62}57 & \cellcolor{prevcol!62}41 \\
Pneumothorax & \cellcolor{prevcol!13}1\textperiodcentered{}7 &  & \cellcolor{prevcol!21}4\textperiodcentered{}2 & \cellcolor{prevcol!8}$<$1 & \cellcolor{prevcol!6}$<$1 & \cellcolor{prevcol!8}$<$1 & \cellcolor{prevcol!28}7\textperiodcentered{}8 & \cellcolor{prevcol!44}19 & \cellcolor{prevcol!20}4\textperiodcentered{}1 & \cellcolor{prevcol!26}7\textperiodcentered{}0 &  & \cellcolor{prevcol!26}7\textperiodcentered{}0 \\
\rowcolor{rowband}Atelectasis & \cellcolor{prevcol!55}31 &  &  & \cellcolor{prevcol!31}9\textperiodcentered{}5 & \cellcolor{prevcol!20}4\textperiodcentered{}1 & \cellcolor{prevcol!17}2\textperiodcentered{}9 & \cellcolor{prevcol!62}56 & \cellcolor{prevcol!62}38 & \cellcolor{prevcol!8}$<$1 &  & \cellcolor{prevcol!62}38 & \cellcolor{prevcol!62}49 \\
Cardiomegaly & \cellcolor{prevcol!56}31 &  &  & \cellcolor{prevcol!26}6\textperiodcentered{}8 & \cellcolor{prevcol!30}8\textperiodcentered{}8 & \cellcolor{prevcol!32}10 & \cellcolor{prevcol!62}54 & \cellcolor{prevcol!37}13 & \cellcolor{prevcol!34}11 &  & \cellcolor{prevcol!62}85 & \cellcolor{prevcol!57}32 \\
\rowcolor{rowband}Lung Opacity & \cellcolor{prevcol!62}53 & \cellcolor{prevcol!62}80 & \cellcolor{prevcol!60}35 & \cellcolor{prevcol!33}11 &  & \cellcolor{prevcol!17}2\textperiodcentered{}8 & \cellcolor{prevcol!62}63 & \cellcolor{prevcol!62}81 &  &  & \cellcolor{prevcol!62}40 & \cellcolor{prevcol!59}35 \\
Consolidation & \cellcolor{prevcol!25}6\textperiodcentered{}2 &  &  &  & \cellcolor{prevcol!9}$<$1 & \cellcolor{prevcol!18}3\textperiodcentered{}2 & \cellcolor{prevcol!48}23 & \cellcolor{prevcol!49}24 & \cellcolor{prevcol!9}$<$1 &  & \cellcolor{prevcol!62}78 & \cellcolor{prevcol!30}9\textperiodcentered{}1 \\
\rowcolor{rowband}Edema & \cellcolor{prevcol!40}16 &  &  & \cellcolor{prevcol!14}2\textperiodcentered{}1 & \cellcolor{prevcol!6}$<$1 &  & \cellcolor{prevcol!62}46 & \cellcolor{prevcol!6}$<$1 & \cellcolor{prevcol!6}$<$1 &  & \cellcolor{prevcol!62}55 & \cellcolor{prevcol!58}33 \\
Pneumonia & \cellcolor{prevcol!16}2\textperiodcentered{}5 &  &  & \cellcolor{prevcol!14}1\textperiodcentered{}9 & \cellcolor{prevcol!18}3\textperiodcentered{}2 &  & \cellcolor{prevcol!55}30 & \cellcolor{prevcol!60}36 & \cellcolor{prevcol!53}28 & \cellcolor{prevcol!49}24 &  & \cellcolor{prevcol!15}2\textperiodcentered{}3 \\
\rowcolor{rowband}Infiltration &  &  &  & \cellcolor{prevcol!17}2\textperiodcentered{}8 & \cellcolor{prevcol!27}7\textperiodcentered{}4 & \cellcolor{prevcol!14}1\textperiodcentered{}9 &  & \cellcolor{prevcol!59}35 & \cellcolor{prevcol!6}$<$1 &  & \cellcolor{prevcol!37}14 & \cellcolor{prevcol!55}30 \\
Lung Lesion & \cellcolor{prevcol!11}1\textperiodcentered{}2 &  &  & \cellcolor{prevcol!6}$<$1 &  & \cellcolor{prevcol!18}3\textperiodcentered{}1 & \cellcolor{prevcol!41}17 & \cellcolor{prevcol!49}24 &  &  & \cellcolor{prevcol!62}90 & \cellcolor{prevcol!29}8\textperiodcentered{}5 \\
\rowcolor{rowband}Enlarged cardiomediastinum & \cellcolor{prevcol!62}49 &  &  &  &  &  & \cellcolor{prevcol!42}18 & \cellcolor{prevcol!23}5\textperiodcentered{}2 & \cellcolor{prevcol!6}$<$1 &  & \cellcolor{prevcol!62}48 & \cellcolor{prevcol!35}12 \\
Pleural Thickening &  &  &  & \cellcolor{prevcol!9}$<$1 & \cellcolor{prevcol!18}3\textperiodcentered{}3 & \cellcolor{prevcol!24}5\textperiodcentered{}6 &  & \cellcolor{prevcol!62}58 & \cellcolor{prevcol!29}8\textperiodcentered{}6 &  &  & \cellcolor{prevcol!49}24 \\
\rowcolor{rowband}Emphysema &  &  &  & \cellcolor{prevcol!16}2\textperiodcentered{}4 & \cellcolor{prevcol!9}$<$1 &  &  & \cellcolor{prevcol!48}23 & \cellcolor{prevcol!12}1\textperiodcentered{}4 &  &  &  \\
Mass &  &  &  & \cellcolor{prevcol!6}$<$1 & \cellcolor{prevcol!9}$<$1 &  &  & \cellcolor{prevcol!34}12 & \cellcolor{prevcol!13}1\textperiodcentered{}6 &  &  &  \\
\rowcolor{rowband}Nodule &  &  &  & \cellcolor{prevcol!14}1\textperiodcentered{}9 & \cellcolor{prevcol!19}3\textperiodcentered{}5 &  &  & \cellcolor{prevcol!53}28 & \cellcolor{prevcol!22}4\textperiodcentered{}8 &  &  &  \\
Nodule or mass &  &  & \cellcolor{prevcol!37}14 &  &  & \cellcolor{prevcol!24}5\textperiodcentered{}9 &  &  &  &  & \cellcolor{prevcol!62}61 & \cellcolor{prevcol!29}8\textperiodcentered{}2 \\
\rowcolor{rowband}Support Devices & \cellcolor{prevcol!62}51 &  &  &  & \cellcolor{prevcol!14}1\textperiodcentered{}8 &  & \cellcolor{prevcol!62}63 &  & \cellcolor{prevcol!15}2\textperiodcentered{}3 &  &  &  \\
Fibrosis &  &  &  & \cellcolor{prevcol!7}$<$1 & \cellcolor{prevcol!7}$<$1 &  &  &  & \cellcolor{prevcol!6}$<$1 &  &  &  \\
\rowcolor{rowband}Hernia &  &  &  & \cellcolor{prevcol!12}1\textperiodcentered{}5 & \cellcolor{prevcol!13}1\textperiodcentered{}6 &  &  &  & \cellcolor{prevcol!6}$<$1 &  &  &  \\
Pleural Other & \cellcolor{prevcol!9}$<$1 &  &  &  &  &  & \cellcolor{prevcol!31}9\textperiodcentered{}6 &  & \cellcolor{prevcol!8}$<$1 &  &  &  \\
\rowcolor{rowband}Aortic Atherosclerosis &  &  &  &  & \cellcolor{prevcol!14}2\textperiodcentered{}1 &  &  &  & \cellcolor{prevcol!62}44 &  &  &  \\
Aortic Enlargement &  &  &  &  &  & \cellcolor{prevcol!27}7\textperiodcentered{}3 &  &  & \cellcolor{prevcol!12}1\textperiodcentered{}4 &  &  &  \\
\rowcolor{rowband}Calcification &  &  &  &  &  & \cellcolor{prevcol!25}6\textperiodcentered{}5 &  &  & \cellcolor{prevcol!17}2\textperiodcentered{}8 &  &  &  \\
Granuloma &  &  &  & \cellcolor{prevcol!32}10 & \cellcolor{prevcol!16}2\textperiodcentered{}5 &  &  &  &  &  &  &  \\
\rowcolor{rowband}ILD &  &  &  &  &  & \cellcolor{prevcol!27}7\textperiodcentered{}4 &  &  & \cellcolor{prevcol!21}4\textperiodcentered{}4 &  &  &  \\
Scoliosis &  &  &  &  & \cellcolor{prevcol!25}6\textperiodcentered{}3 &  &  &  & \cellcolor{prevcol!12}1\textperiodcentered{}3 &  &  &  \\
\bottomrule
\end{tabular}%
\end{adjustbox}

\smallskip
\begin{minipage}{\textwidth}\scriptsize
\emph{Evaluated, $n$} is the cohort scored in this study; \emph{per-task prevalence} is the percentage of that cohort positive for the label (blank = label not provided by the dataset; darker shading = more prevalent; ``$<$1'' = non-zero but below 1\%). Full definitions of every row, the label-merging rules, and the cohort-specific notes are given in the appendix, which lists the same columns over all 52 tasks. TAIX~Ray was publicly released in February 2026, after every evaluated model had been built; it is grouped with the private cohorts because no evaluated model can have been trained on it.
\end{minipage}
\end{table}

%% file: tables/table2_models.tex
\begin{table}[p]
\centering
\small
\setlength{\tabcolsep}{4pt}
\renewcommand{\arraystretch}{1.25}
\caption{The ten evaluated chest X-ray diagnostic models, grouped by the four
learning paradigms, with the architecture, the data each was trained on, the
training objective, the native inference interface, and the source reference.
Checkpoints and variants are those used in our evaluation. MIMIC-CXR comprises
$\sim$377k images / $\sim$227k studies with free-text reports. Within the
vision-language group the training strategies differ---contrastive (CheXzero,
BioViL-T), knowledge-enhanced alignment (MedKLIP, MAVL), and prompt learning
(PsPG)---as reflected in the objective column. CheXagent returns categorical
Yes/No answers rather than calibrated scores and is therefore excluded from the
probability-based operating-point analyses.}
\label{tab:models}
\begin{adjustbox}{max width=\textwidth,max totalheight=0.86\textheight,center}
\begin{tabular}{L{2.0cm} L{3.0cm} L{3.9cm} L{4.4cm} L{3.2cm} L{2.4cm}}
\toprule
\textbf{Model} & \textbf{Architecture} & \textbf{Training data} &
\textbf{Training paradigm / objective} & \textbf{Inference interface} &
\textbf{Reference} \\
\midrule

\multicolumn{6}{l}{\textit{Supervised CNN classifiers}}\\
\addlinespace[2pt]
X-Raydar
  & XNet38MS: multi-scale Inception-v3 CNN ensemble (299/512/1024\,px)
  & 2,513,546 CXR studies from six UK hospitals across three NHS Trusts (2006--2019); labels NLP-extracted from reports
  & Supervised multi-label classification of 37 radiological findings
  & Image $\rightarrow$ 37 finding probabilities (scale-averaged)
  & Cid et al.\citep{cid2024xraydar} \\
\addlinespace[3pt]
DenseNet (TorchXRay\-Vision)
  & DenseNet-121 CNN (res 224)
  & ``All'' checkpoint: union of public CXR sets---NIH ChestX-ray14, CheXpert, MIMIC-CXR, PadChest, RSNA Pneumonia, OpenI---with harmonized labels
  & Supervised multi-label classification (per-label BCE)
  & Image $\rightarrow$ 18 label logits (probabilities)
  & Cohen et al.\citep{cohen2022torchxrayvision}; Huang et al.\citep{huang2017densenet} \\
\addlinespace[3pt]
ResNet (TorchXRay\-Vision)
  & ResNet-50 CNN (res 512)
  & Same multi-dataset union (``all'' checkpoint)
  & Supervised multi-label classification (per-label BCE)
  & Image $\rightarrow$ 18 label logits (probabilities)
  & Cohen et al.\citep{cohen2022torchxrayvision}; He et al.\citep{he2016resnet} \\
\addlinespace[4pt]
\midrule

\multicolumn{6}{l}{\textit{Vision-language foundation models}}\\
\addlinespace[2pt]
CheXzero
  & CLIP: ViT-B/32 image encoder + Transformer text encoder
  & MIMIC-CXR image--report pairs; no explicit disease labels
  & Self-supervised image--text contrastive learning (CLIP/InfoNCE)
  & Zero-shot text-prompt similarity $\rightarrow$ probability
  & Tiu et al.\citep{tiu2022chexzero} \\
\addlinespace[3pt]
BioViL-T
  & Hybrid CNN--Transformer multi-image encoder + CXR-BERT
  & MIMIC-CXR, using temporally linked image sequences and reports
  & Temporal vision-language pre-training; global + local contrastive objectives
  & Zero-shot text-prompt similarity $\rightarrow$ probability
  & Bannur et al.\citep{bannur2023biovilt} \\
\addlinespace[3pt]
MedKLIP
  & ResNet-50 image encoder + BERT entity encoder + Transformer fusion
  & MIMIC-CXR reports distilled into (entity, position, existence) triplets, queried against a medical knowledge base
  & Knowledge-enhanced pre-training supervising entity existence and location
  & Zero-shot entity-description query $\rightarrow$ probability
  & Wu et al.\citep{wu2023medklip} \\
\addlinespace[3pt]
MAVL
  & ResNet-50 image + BERT text + Transformer fusion (MedKLIP family)
  & MIMIC-CXR with LLM-generated multi-aspect disease descriptions
  & Knowledge-enhanced pre-training; diseases decomposed into interpretable visual aspects
  & Zero-shot multi-aspect description similarity $\rightarrow$ probability
  & Phan et al.\citep{phan2024mavl} \\
\addlinespace[3pt]
PsPG
  & Pretrained CLIP (RN50) + RNN pseudo-prompt decoder
  & Multi-label CXR datasets (CheXpert / MIMIC-CXR family), adapted on top of the pretrained CLIP backbone
  & Prompt learning; autoregressive class-tailored pseudo-prompts (spatial fusion, soft co-occurrence loss)
  & Generated prompt vs.\ image embedding $\rightarrow$ per-label probability
  & Ye et al.\citep{ye2024pspg} \\
\addlinespace[4pt]
\midrule

\multicolumn{6}{l}{\textit{Instruction-tuned QA foundation model}}\\
\addlinespace[2pt]
CheXagent
  & $\sim$8B VLM: CXR vision encoder + bridge + Mistral-7B LLM
  & CheXinstruct: $\sim$6M image--question--answer triplets aggregated from $\sim$28 public CXR datasets
  & Instruction tuning (generative, multi-task)
  & Image + prompt $\rightarrow$ free text, parsed to a Yes/No label
  & Chen et al.\citep{chen2024chexagent} \\
\addlinespace[4pt]
\midrule

\multicolumn{6}{l}{\textit{Repurposed foundation-model encoder}}\\
\addlinespace[2pt]
CheXagent-AutoGluon
  & Frozen CheXagent-8b image embeddings + AutoGluon tabular ensemble head
  & CheXpert training set (225,008 radiographs); head trained on 14 findings plus an ``Any Abnormality'' aggregate (appendix section 5)
  & Supervised stacked ensemble on image embeddings (in-house)
  & Image $\rightarrow$ embedding $\rightarrow$ per-label probability
  & This study; Chen et al.\citep{chen2024chexagent}; Erickson et al.\citep{erickson2020autogluon} \\
\bottomrule
\end{tabular}
\end{adjustbox}
\end{table}

%% file: sections/03_results.tex
\section{Results}\label{sec:results}

After harmonizing labels across 12 datasets and 10 models as described in Methods, we evaluated model performance using specificity at 95\% sensitivity as the primary clinically oriented operating-point metric. Across all available dataset-task units, MAVL and MedKLIP achieved the strongest overall specificity at 95\% sensitivity, followed by DenseNet and ResNet. CheXagent-AutoGluon showed intermediate performance, whereas CheXzero, BioViL-T, PsPG, and X-Raydar showed lower operating-point specificity.

Despite this relative ranking, absolute performance at the high-sensitivity threshold remained modest across all models. The best overall specificity at 95\% sensitivity was approximately 0\textperiodcentered{}37 (MAVL 0\textperiodcentered{}37, 95\% CI 0\textperiodcentered{}31--0\textperiodcentered{}44; MedKLIP 0\textperiodcentered{}37, 95\% CI 0\textperiodcentered{}30--0\textperiodcentered{}44). Clinically, this means that when the model was calibrated to detect approximately 95\% of positive examinations, only about one-third of truly negative examinations could be correctly cleared. At the same operating point, PPV was approximately 0\textperiodcentered{}19 for the leading models (MAVL 0\textperiodcentered{}18, 95\% CI 0\textperiodcentered{}13--0\textperiodcentered{}24; MedKLIP 0\textperiodcentered{}19, 95\% CI 0\textperiodcentered{}13--0\textperiodcentered{}25), indicating that most flagged examinations would be false positives. These findings suggest that current models remain limited as standalone high-sensitivity screening or rule-out tools.

\begin{figure}[tbp]
    \centering
    \includegraphics[width=\linewidth]{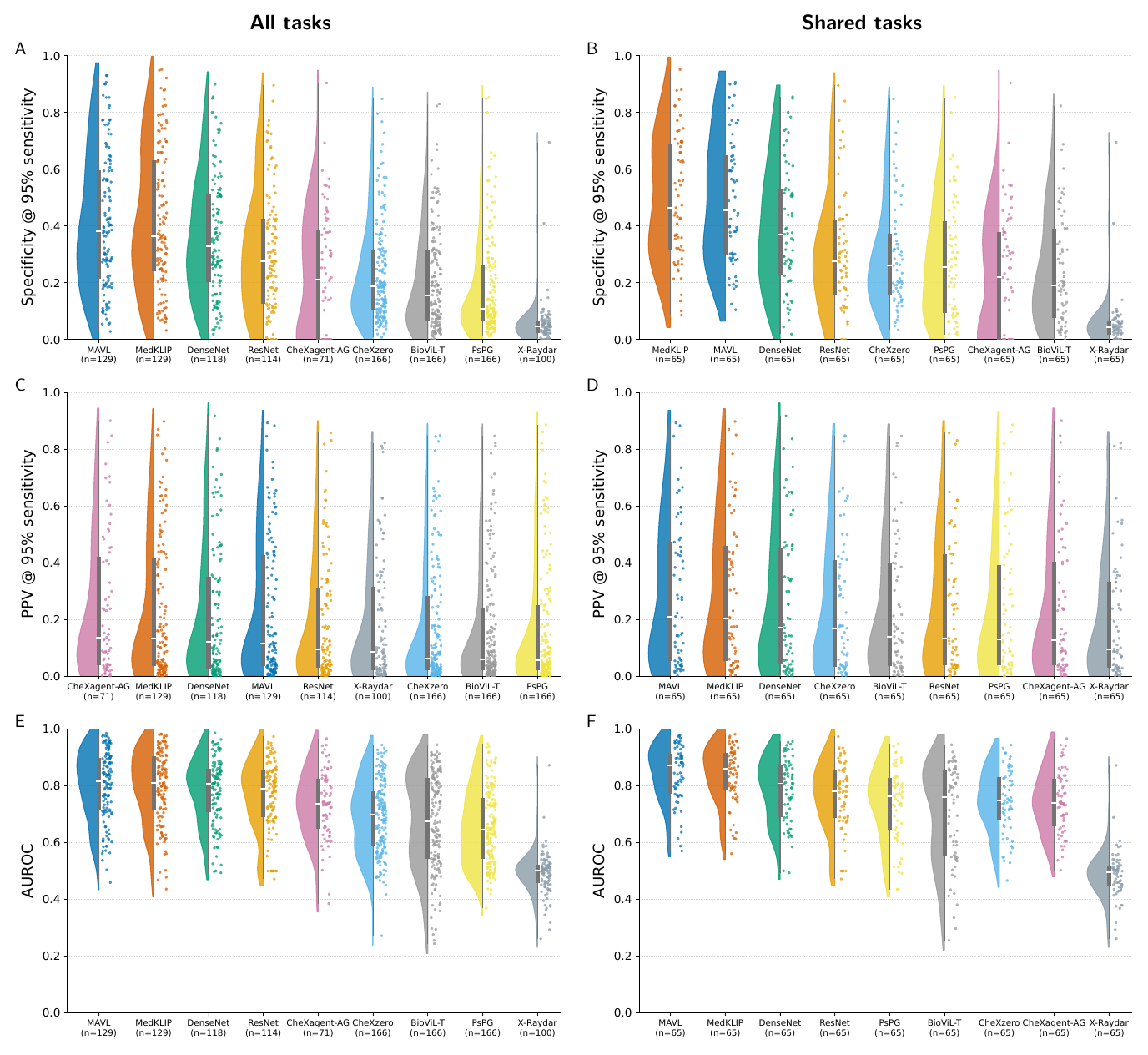}
    \caption{\textbf{Model ranking across all units and shared tasks.} Raincloud distributions of per-unit performance for each model, with models ordered by median within each panel. Rows correspond to the three metrics---specificity at 95\% sensitivity (\textbf{A}, \textbf{B}), PPV at 95\% sensitivity (\textbf{C}, \textbf{D}), and AUROC (\textbf{E}, \textbf{F}); the left column (\textbf{A}, \textbf{C}, \textbf{E}) covers all available dataset-task units and the right column (\textbf{B}, \textbf{D}, \textbf{F}) the fair shared-task set available across all probabilistic models. MAVL and MedKLIP achieved the strongest operating-point performance overall and retained leading performance in the shared-task analysis. AUROC showed less clinically informative separation, emphasizing the importance of operating-point evaluation for screening and rule-out applications.}
    \label{fig:ranking}
\end{figure}

Because the number of evaluable dataset-task units differed across models, we also performed a matched shared-task analysis. In this matched comparison, MAVL and MedKLIP retained leading performance, confirming that their overall ranking was not solely driven by differences in task coverage. However, the per-unit distributions remained broad for every model, indicating substantial variability across individual dataset-task combinations (\cref{fig:ranking}).

We next compared model performance between public benchmark datasets and private clinical cohorts. On public benchmarks, MAVL, DenseNet, and MedKLIP achieved the highest specificity at 95\% sensitivity, with leading values around 0\textperiodcentered{}42--0\textperiodcentered{}43. On private cohorts, specificity decreased for most models. MedKLIP achieved the highest private-cohort specificity at approximately 0\textperiodcentered{}33, followed by MAVL, ResNet, and DenseNet.

\begin{figure}[tbp]
    \centering
    \includegraphics[width=\linewidth]{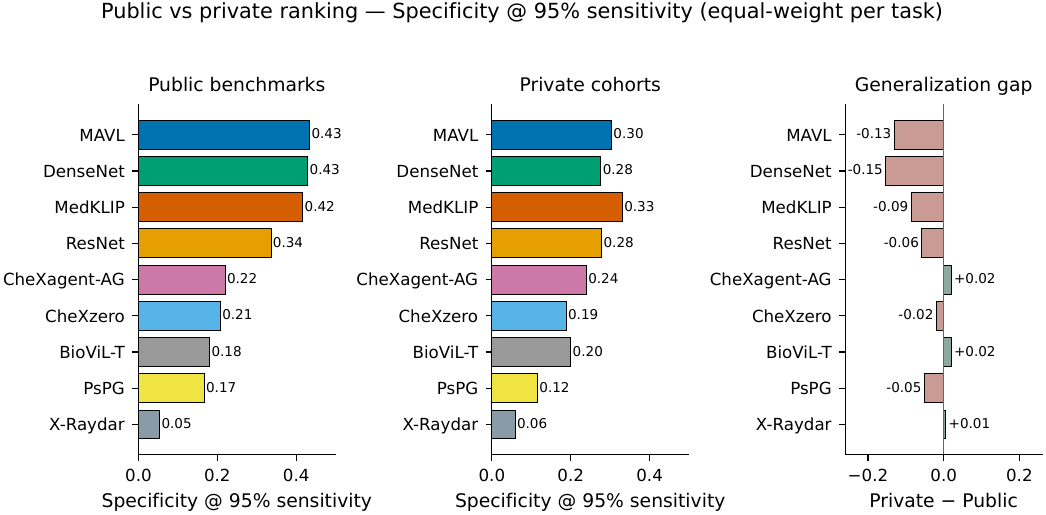}
    \caption{\textbf{Public versus private generalization.} Specificity at 95\% sensitivity on public benchmarks and private cohorts, with the generalization gap calculated as private minus public performance. Most models showed lower specificity on private data than on public benchmarks. MedKLIP achieved the highest private-cohort specificity, whereas MAVL and DenseNet showed larger public-private decreases. These results indicate that public benchmark performance can overestimate real-world rule-out performance.}
    \label{fig:public-private}
\end{figure}

The public-private generalization gap was negative for several leading models. MAVL showed an approximate decrease of 0\textperiodcentered{}13 from public to private datasets, and DenseNet showed an approximate decrease of 0\textperiodcentered{}15. MedKLIP had a smaller but still negative gap of approximately 0\textperiodcentered{}09 and was the strongest model on private cohorts. These results indicate that public benchmarks may overestimate real-world rule-out performance and that private external validation is essential before clinical deployment (\cref{fig:public-private}).

Task-level analysis revealed pronounced heterogeneity across diagnostic findings. MAVL and MedKLIP performed strongly across several pulmonary and pleural findings, including edema, cardiomegaly, pleural effusion, consolidation, and pneumothorax. DenseNet was competitive for selected structural or diffuse findings, whereas CheXagent-AutoGluon showed isolated strength for selected tasks, including device-related classification.

\begin{figure}[tbp]
    \centering
    \includegraphics[width=\linewidth]{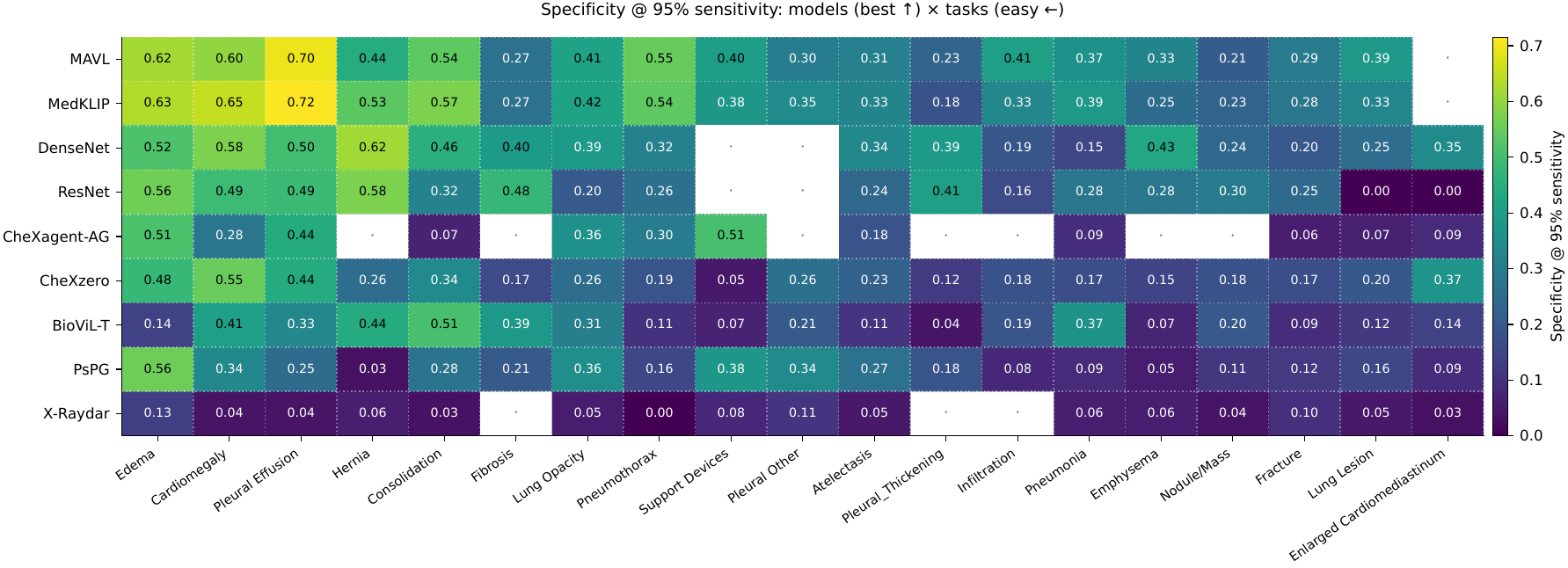}
    \caption{\textbf{Task-specific model heterogeneity.} Model-by-task heatmap of specificity at 95\% sensitivity. Rows represent models ordered by overall performance, and columns represent diagnostic tasks ordered by average difficulty. No single model dominated all findings, demonstrating substantial task-level heterogeneity and supporting task-aware benchmarking.}
    \label{fig:heatmap}
\end{figure}

No single model dominated all diagnostic findings. Per-task winners varied across categories, demonstrating that aggregate model ranking can obscure clinically important task-specific strengths and failures. These results support task-aware benchmarking rather than reliance on a single overall leaderboard (\cref{fig:heatmap}).

\begin{figure}[p]
    \centering
    \includegraphics[width=\linewidth,height=0.77\textheight,keepaspectratio]{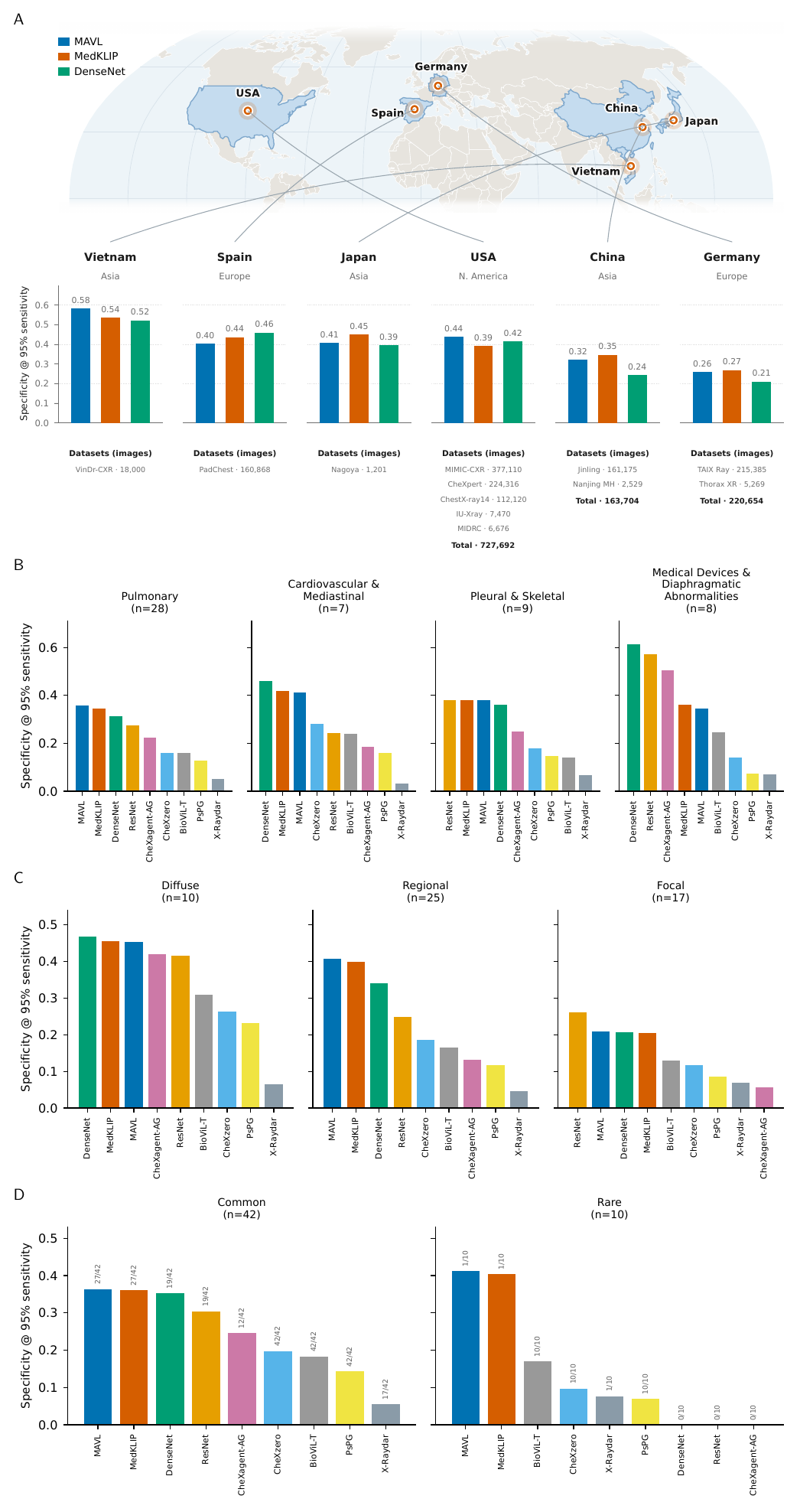}
    \caption{\textbf{Country-level, diagnostic-category, spatial-extent, and prevalence-related heterogeneity in model performance.} \textbf{A}, Country-level specificity at 95\% sensitivity across six countries, ordered from the best- to the worst-performing country, for the three models with the highest specificity when data were pooled across all cohorts; the same three models are shown for every country, so panel A displays their country-to-country ordering rather than the top three models within each country. \textbf{B}, Specificity at 95\% sensitivity by diagnostic category. \textbf{C}, Specificity at 95\% sensitivity by spatial extent, showing lower performance for focal findings than for regional or diffuse findings in seven of nine models. \textbf{D}, Specificity at 95\% sensitivity by prevalence, showing reduced performance for rare findings. Across panels, model ranking varied by clinical and dataset context, indicating that no single model was uniformly optimal across all strata.}
    \label{fig:strata}
\end{figure}

When performance was stratified by country, the relative ordering of the leading models differed from country to country (\cref{fig:strata}A). MAVL, MedKLIP, and DenseNet---the three strongest models when data were pooled across all cohorts, and the models shown in \cref{fig:strata}A---were together the top three in four of the six countries. They were not in Spain, where ResNet reached the highest country-level specificity at 95\% sensitivity (0\textperiodcentered{}48) and MAVL ranked fourth (0\textperiodcentered{}40), nor in Germany, where DenseNet ranked fourth. Leading models therefore remained broadly competitive across countries, while absolute performance and rank order remained sensitive to population differences, acquisition protocols, label definitions, and institutional practice patterns.

Performance varied not only with where the data originated but also with what was being detected. Model rankings shifted across pulmonary, cardiovascular or mediastinal, pleural or skeletal, device-related, and diaphragmatic abnormalities (\cref{fig:strata}B). Stratified by the spatial extent of the finding, specificity at 95\% sensitivity was lowest for focal findings (mean across the nine probabilistic models 0\textperiodcentered{}15), intermediate for regional findings (0\textperiodcentered{}23), and highest for diffuse findings (0\textperiodcentered{}34; \cref{fig:strata}C). This ordering held for seven of nine models; ResNet showed slightly higher specificity on focal than on regional findings, and X-Raydar showed a fully inverted ordering, although its specificity was uniformly low (0\textperiodcentered{}05--0\textperiodcentered{}07) across all three strata. For the three leading models, specificity fell from 0\textperiodcentered{}45--0\textperiodcentered{}47 on diffuse findings to approximately 0\textperiodcentered{}21 on focal findings. Focal abnormalities were challenging for every model, likely reflecting their smaller spatial footprint and greater visual subtlety, whereas diffuse abnormalities produced broader image-level changes that were more readily detected.

Prevalence-stratified analysis showed that rare findings remained challenging for most models. MAVL and MedKLIP maintained relatively stronger performance on rare findings than other models, suggesting that knowledge-enhanced vision-language models may better handle low-prevalence abnormalities. However, absolute operating-point performance for rare findings remained limited, highlighting the need to evaluate model robustness on long-tail clinical tasks (\cref{fig:strata}D).

%% file: sections/04_discussion.tex
\section{Discussion}\label{sec:discussion}

In this multinational benchmark of 10 CXR classification models across public and private datasets, we found that current models remain limited as standalone high-sensitivity rule-out tools. MAVL and MedKLIP were the most consistent overall performers, but the best specificity at 95\% sensitivity was only approximately 0\textperiodcentered{}37. At the same operating point, PPV remained low, suggesting that most flagged examinations would be false positives. These findings indicate that relative model ranking should be interpreted separately from clinical usability.

These results extend a small body of independent head-to-head evaluations. The largest published comparison scored 12 commercial computer-aided detection products, but in 774 people from one country and for one target condition;\citep{qin2024cadsouthafrica} the most geographically diverse compared seven products across seven countries, again for tuberculosis alone and at 90\% rather than 95\% sensitivity.\citep{worodria2026multicountry} Both evaluated commercial detectors for a single disease, on 774 and 3901 people respectively; we evaluate published research and vision-language models across a broad multi-label finding taxonomy, on 466\,135 radiographs---nearly 20 times as many as the largest previous head-to-head comparison---and report positive predictive value alongside specificity at the operating point, which no previous study has done for more than one model.

Our study emphasizes the value of evaluating CXR AI at clinically relevant operating points. In triage, screening, and rule-out applications, the clinical threshold is often chosen to prioritize sensitivity. Once sensitivity is fixed, specificity determines how much work is actually saved. A model with modest specificity at 95\% sensitivity may be safe in the sense that it misses relatively few abnormalities, but it may still offer limited workflow benefits because it flags many negative examinations. This distinction is critical for translating benchmark results into clinical deployment decisions.

We also found that public benchmarks overestimated private-cohort performance for most models. Private clinical cohorts may differ in acquisition protocol, patient population, disease prevalence, image quality, label definitions, and reporting practice. The observed public-private performance gap suggests that model evaluation on public datasets alone is insufficient for estimating real-world performance.\citep{zech2018confounding} Prospective and multi-site validation across private clinical environments remains necessary before deployment.

Knowledge-enhanced vision-language models showed the strongest overall performance. MAVL and MedKLIP were consistently among the leading models across overall, shared-task, and country-level analyses. Their relative strength may reflect the value of structured medical knowledge, prompt design, or image-text alignment for radiographic abnormality recognition. However, these models did not dominate every task. DenseNet and other supervised models remained competitive for selected categories, and CheXagent-AutoGluon showed isolated strengths.

Task heterogeneity was substantial: no model was uniformly superior across diagnostic findings, focal abnormalities were the hardest for every model, and rare findings remained difficult, exposing limitations in robustness for long-tail clinical abnormalities.

Our results have several implications for future CXR AI development. First, operating-point-aware development and calibration should be prioritized over AUROC-only leaderboards. Second, public benchmarks should be supplemented with private, geographically diverse external validation. Third, uncertainty-aware triage methods, including conformal prediction,\citep{angelopoulos2021conformal} may help identify cases where model predictions are unreliable.

This study has several limitations. First, the benchmark covers a single task type. All 10 models were evaluated as image-level classifiers of radiographic findings; segmentation, lesion localization, and report generation were not assessed, and the one evaluated model able to generate free text was scored by parsing its answer into a yes/no label. The conclusions drawn here therefore apply to CXR classification and should not be read as an assessment of CXR AI across all task types. Second, retrospective benchmark evaluation may not fully capture prospective clinical workflow performance. Third, label harmonization across datasets is inherently imperfect because source datasets differ in label definitions, annotation processes, and reporting conventions. Fourth, not all models produced comparable probability outputs for all tasks. In particular, QA-based models such as CheXagent may require different evaluation strategies from probability-based classifiers. Fifth, PPV depends strongly on disease prevalence and should be interpreted in relation to cohort composition. Sixth, although private cohorts improved external validity, additional institutions, scanners, countries, and prospective settings are needed to assess deployment readiness; one of those cohorts, TAIX Ray, has since been released publicly, although only after every evaluated model had been built. Finally, operating-point thresholds estimated retrospectively may require recalibration before clinical use.

In summary, across 12 datasets from six countries, current CXR classification models fall short of what a high-sensitivity rule-out workflow would require: at 95\% sensitivity the best of them clears only about a third of truly negative examinations, and most flagged examinations would be false positives. Performance varied substantially by public versus private data source, country, diagnostic task, spatial extent, and prevalence, so a model's position on a single overall leaderboard is not a sufficient basis for deployment. Operating-point-aware calibration, external validation on institutional cohorts, and uncertainty-aware deployment strategies are needed before these models can be relied on for high-sensitivity screening or rule-out.

%% file: sections/06_statements.tex
\section*{Contributors}

QX and YiL contributed equally to this work. QX and YiL conceived and designed the study, implemented the evaluation framework, performed the analyses, interpreted the results, and drafted the manuscript. XZ, BD, and MA provided guidance on the study design and evaluation strategy. AGE curated and prepared the public benchmark datasets for analysis. WZ, SS, RH, MI, and AH curated and provided the Nagoya cohort. ZZ, MvdS, MG, and DT curated and provided the Thorax XR and TAIX Ray cohorts. CX, SL, MY, YaL, ZS, and CZ curated and provided the Chinese clinical cohorts, including the Jinling and Nanjing Multi-Hospital cohorts. HI, XL, and OG supervised the study. XL and OG acquired funding. The underlying data were directly accessed and verified at the level of the individual radiograph, covering both the source images and the reference labels, by AGE, QX, and YiL for the seven public benchmarks; by WZ, AH, SS, RH, and MI for the Nagoya cohort; by ZZ, MvdS, MG, and DT for the Thorax XR and TAIX Ray cohorts; by CX and QX for the Nanjing Multi-Hospital cohort; and by SL, MY, YaL, ZS, CZ, and XL for the Jinling cohort. QX and YiL directly accessed and verified the pooled study dataset, the cross-cohort label mapping, and the image-level model outputs from which every reported number and figure was computed. All authors had access to the aggregated study data, reviewed and edited the manuscript, approved the final version, and accept responsibility for the decision to submit for publication.

\section*{Declaration of interests}

DT holds shares in StratifAI and Synagen, and has received honoraria from Bayer, AstraZeneca, Philips, Roche, Pfizer, and Gilead. All other authors declare no competing interests.

\section*{Data sharing}

No individual participant data are shared. The five institutional cohorts (Jinling, Nanjing Multi-Hospital, Nagoya, Thorax XR, and TAIX Ray) contain clinical imaging data governed by institutional and national data protection restrictions and cannot be redistributed by the authors; requests for access should be directed to the corresponding authors and are subject to institutional review and a data use agreement. TAIX Ray has since been released independently by its originating institution under a CC~BY~4.0 licence.\cite{truhn2026taixray}

The seven public benchmarks are available from their original custodians: CheXpert (\url{https://stanfordmlgroup.github.io/competitions/chexpert/}), MIMIC-CXR (\url{https://physionet.org/content/mimic-cxr/}), ChestX-ray14 (\url{https://nihcc.app.box.com/v/ChestXray-NIHCC}), MIDRC (\url{https://data.midrc.org/}), IU-Xray (\url{https://openi.nlm.nih.gov/}), PadChest (\url{https://bimcv.cipf.es/bimcv-projects/padchest/}), and VinDr-CXR (\url{https://physionet.org/content/vindr-cxr/}). MIMIC-CXR and VinDr-CXR require PhysioNet credentialling and a signed data use agreement.

All de-identified derived data supporting the reported analyses are available now, without restriction and without an end date, to any user and for any purpose. The complete performance-metrics table from which every reported number and figure is generated---36\,885 rows giving all nine metrics for every model, dataset, diagnostic task, and age stratum, each with its bootstrap 95\% CI, standard error, and positive and negative sample counts---is distributed in Apache Parquet and CSV format, with a definition of every column, alongside the analysis code at \url{https://github.com/gevaertlab/chest-xray-benchmark} (Apache License 2.0). That repository also contains the evaluation framework, the label harmonisation logic, the metric computation with bootstrap CIs, and the figure-generation code, so that every figure and table in this Article can be regenerated from a clone with a single command, without a GPU and without access to any of the 12 datasets. It contains no patient-level or per-image data. The study protocol was not registered and no separate statistical analysis plan document exists; the analysis is specified in full in the Methods and in the appendix. Pretrained weights for the 10 evaluated models were obtained from their original public releases as cited in the Methods and used without modification, except for the CheXagent-AutoGluon variant, whose training configuration is provided in the repository and described in the appendix.

\section*{Declaration of generative AI use in the writing process}

During the preparation of this work the authors used Claude (Opus 5, Anthropic) to check English grammar and spelling. After using this tool the authors reviewed and edited the content as needed and take full responsibility for the content of the published article. No artificial intelligence tool was used to design the study, to generate, analyse, or interpret any result, or to produce any figure or table.

%% file: sections/S0_cohort_selection.tex
\section{Private cohort selection and image inclusion}\label{sec:supp-cohort-selection}

The five private clinical cohorts---Nagoya, Jinling, Nanjing Multi-Hospital, Thorax XR, and TAIX Ray---were retrospectively assembled from the radiology archives of the participating institutions. TAIX Ray has since been released publicly, after all evaluated models were built; it is analysed with the private cohorts for the reason given in the Methods. For each cohort, all consecutive chest radiographs meeting the eligibility criteria during the prespecified study period were retrospectively included. The Nagoya cohort comprised examinations acquired between January 25, 2016, and June 26, 2025; the Jinling cohort, between January 1, 2017, and December 1, 2025; the Nanjing Multi-Hospital cohort, between August 6, 2017, and September 14, 2024; the TAIX Ray cohort, between January 1, 2010, and December 1, 2023; and the Thorax XR cohort, between January 1, 2010, and April 1, 2024. Each cohort underwent institution-specific eligibility screening and de-identification before inclusion in the benchmark.

Eligible images were frontal anteroposterior or posteroanterior chest radiographs with a usable reference label for at least one evaluated diagnostic finding. Examinations outside the prespecified study periods, non-frontal images when projection information was available, unreadable or corrupted image files, duplicate examinations where applicable, and examinations without usable reference labels were excluded during local cohort curation.

Following eligibility screening and de-identification, 385{,}559 images from the private cohorts were included in the benchmark: 1{,}201 from Nagoya, 161{,}175 from Jinling, 2{,}529 from the Nanjing Multi-Hospital cohort, 215{,}385 from TAIX Ray, and 5{,}269 from Thorax XR. The included images subsequently underwent label harmonization and model inference. Performance was evaluated at the dataset--label--model level. Dataset--label--model units were excluded from the corresponding operating-point analysis if they contained no positive examinations, no negative examinations, or no available model output. The remaining units were included in the coverage-aware analysis, and units available across all probabilistic models were additionally included in the shared-task analysis.

%% file: sections/S0b_reference_standard.tex
\section{Reference-standard ascertainment for the institutional cohorts}\label{sec:supp-reference-standard}

Reference-standard ascertainment was performed locally and was completed before benchmark model inference. Because the five institutional cohorts (\cref{sec:supp-cohort-selection}) were assembled from different clinical reporting environments, the source and ascertainment procedure for the reference labels were cohort-specific. Labels were finalized independently of the evaluated models, and no model output was used to create, select, revise, or adjudicate a source reference label.

For the Nagoya cohort, eligible frontal chest radiographs were retrospectively assessed by local radiologists together with the corresponding finalized clinical radiology reports. Finding-level labels were entered into a structured study annotation sheet and categorized as present, absent, or indeterminate. Examinations considered ambiguous during the initial assessment were reviewed with a senior radiologist before the reference labels were finalized. Indeterminate labels were treated as missing and excluded on a finding-specific basis. Free-text entries recorded under ``other abnormality'' were retained only when they could be mapped unambiguously to a prespecified canonical finding; otherwise, they were treated as unavailable.

For the Jinling cohort, reference labels were obtained from pre-existing structured diagnostic fields associated with routine Chinese-language radiology reports and were curated locally by the institutional radiology team. Because of the size of the cohort, a dedicated independent re-reading of every radiograph was not performed. The source schema contained 41 Chinese-named diagnostic fields, which were translated and mapped to the standardized finding vocabulary as described in \cref{sec:supp-harmonization}. Explicit positive and negative source values were retained. Uncertain, unknown, unavailable, or unmappable values were coded as missing. Two clinically overlapping pairs of source fields were combined using a logical OR, while three source fields without a valid canonical equivalent were excluded from evaluation. Model inference for this cohort was performed locally, and only de-identified reference labels and per-image model outputs were transferred for centralized analysis.

For the Nanjing Multi-Hospital cohort, finding-level reference labels were assigned retrospectively by local radiologists using the frontal chest radiograph and the corresponding finalized clinical report. The same prespecified finding definitions were applied across the participating hospitals. Cases considered ambiguous during local review underwent senior radiologist review before the labels were finalized and before the three institutional subsets were pooled. Findings were recorded as present, absent, or indeterminate. Unresolved or indeterminate assessments were treated as missing rather than classified as negative.

For TAIX Ray, reference labels were taken directly from structured, itemized reports completed during routine clinical interpretation of bedside chest radiographs. The standardized reporting template assessed heart size, pulmonary congestion, left- and right-sided pleural effusion, left- and right-sided pulmonary opacities, and left- and right-sided atelectasis using ordinal categories. Reports were generated by 134 trained radiologists. Each radiograph was interpreted by a single routine clinical reader, and reports produced by readers without completed specialist or board-equivalent certification were validated by a board-certified radiologist. For the present benchmark, the ordinal categories were converted to binary finding-level labels according to the prespecified rules described in \cref{sec:supp-harmonization}. Left- and right-sided labels were combined using a logical OR, and equivocal, non-assessable, or otherwise unusable values were treated as missing.

For the Thorax XR cohort, reference labels were derived from finalized routine clinical radiology reports and converted locally into structured finding-level fields. The finalized report served as the primary clinical reference standard, and the structured labels were curated by the local radiology team before model evaluation. Because the cohort was based on routine clinical reporting, a uniform independent second reading was not available for every examination. Source values coded as ``not given,'' ``unknown,'' indeterminate, unavailable, or otherwise unmappable were treated as missing and excluded on a finding-specific basis.

Across all five cohorts, only explicit source annotations were used. Findings not included in a cohort's native label schema were considered unavailable rather than negative. When a source schema included an explicit study-level ``no finding'' designation, all findings covered by that schema were coded as absent. Duplicate image exports were removed where identifiable, and reference labels were assigned at the radiograph or examination level. Label harmonization was performed only after the cohort-specific source labels had been finalized. Uncertain and missing values were handled separately for each finding, so an examination could remain evaluable for one finding while being excluded for another. Because the cohorts used heterogeneous local ascertainment procedures and several relied on routine single-reader clinical reports, a uniform benchmark-wide inter-reader agreement statistic was not available.

%% file: sections/S1_harmonization.tex
\section{Harmonization of heterogeneous dataset labels}\label{sec:supp-harmonization}

The evaluation datasets were assembled from sources that annotate chest radiographs using incompatible label schemas, differing in language (English, Japanese, Chinese, and Spanish-derived), granularity (finding-level versus radiographic-sign-level), naming convention (for example \texttt{Pleural Effusion} versus \texttt{Effusion}, or \texttt{LungOpacity} versus \texttt{Lung Opacity}), and encoding (binary indicators, ordinal severity grades, and free text). To enable model comparison across datasets, each source's native labels were mapped to a single standardized finding vocabulary of 44 canonical names. The mapping was implemented at the data-loader level so that harmonization is deterministic and reproducible for every cohort, and it was kept intentionally conservative: the large majority of entries are one-to-one renames that only reconcile capitalization, spacing, or minor lexical variants, and the canonical vocabulary deliberately retains near-synonym surface forms (for example \emph{Effusion} and \emph{Pleural Effusion}) rather than forcing an early collapse that could obscure provenance. Uncertain, unknown, or unmappable annotations were uniformly recoded to missing and excluded from scoring rather than being imputed to a class, so that ambiguity in the source never inflates or deflates a metric. Where a dataset marked ``no finding'' at the study level, all findings for that study were set to absent.

A minority of mappings are semantic or lossy and are flagged as such. For the TAIX-Ray cohort, ordinal severity grades (0--4) were thresholded to binary presence (grades 0--1 absent, 2 and 4 present) with the equivocal grade 3 dropped to missing, and left/right laterality columns were combined by a logical OR; its \texttt{PulmonaryCongestion} column was aligned to \emph{Edema}, and the Nagoya \texttt{Cardiomediastinum} column to \emph{Enlarged Cardiomediastinum}. The 161{,}000-image Jinling cohort required the most extensive mapping: 41 Chinese-named label columns were translated to 36 canonical English findings, with two pairs of columns OR-merged and three columns without an English counterpart dropped. Labels for the publicly released datasets (CheXpert, PadChest, IU-Xray, VinDr-CXR, and the ChestX-ray14/Google re-read set) inherit the harmonization already defined in the \texttt{torchxrayvision} library, including its Spanish- and text-derived synonym tables.

Finally, because zero-shot and report-generation models emit heterogeneous output heads, a second, prediction-side alignment maps each model's concept or radiographic-sign outputs onto the canonical findings, aggregating multiple heads by their maximum, and a small canonical-name alias table reconciles residual naming differences at metric-computation time. This prediction-side step operates only on model outputs---ground-truth labels are always read from the true label column---so it can recover otherwise-unscored findings but can never alter a label. The non-identity label-side mappings are enumerated in \cref{tab:label-harmonization}, and the prediction-side alignment in \cref{tab:pred-harmonization}; identity renames that differ only in case or spacing are omitted for brevity.

\input{tables/supS_label_mapping}

Label coverage after harmonization is uneven across cohorts: of the 52 canonical diagnostic tasks that at least one dataset annotates, 27 are annotated by two or more cohorts and 25 by exactly one. Because only tasks shared by several cohorts support cross-dataset comparison, table 1 of the Article lists the 27 multi-cohort tasks. \Cref{tab:dataset-overview-full} reproduces the same table over the full label vocabulary, adding the single-cohort tasks---chiefly the fine-grained Jinling and PadChest findings and the MIDRC COVID-19 pneumonia label---so that the complete coverage and prevalence of every evaluated cohort remains available.

\input{tables/supS_dataset_overview_full}

%% file: tables/supS_label_mapping.tex
\begingroup\small
\setlength{\tabcolsep}{5pt}
\renewcommand{\arraystretch}{1.15}
\begin{longtable}{@{}p{2.4cm}p{4.1cm}p{3.6cm}p{3.1cm}@{}}
\caption{Harmonization of heterogeneous dataset labels to the standardized
finding vocabulary (label side). Only non-identity mappings are shown; renames
that differ solely in capitalization or spacing are omitted. ``OR-merge''
denotes a logical OR across the listed source columns; ``recode'' denotes a
numeric or string re-encoding to binary presence with uncertain/unknown values
set to missing.}
\label{tab:label-harmonization}\\
\toprule
\textbf{Dataset} & \textbf{Original label / column} & \textbf{Canonical finding} & \textbf{Mapping type} \\
\midrule
\endfirsthead
\multicolumn{4}{@{}l}{\emph{Table~\ref{tab:label-harmonization} (continued)}}\\
\toprule
\textbf{Dataset} & \textbf{Original label / column} & \textbf{Canonical finding} & \textbf{Mapping type} \\
\midrule
\endhead
\midrule
\multicolumn{4}{r@{}}{\emph{continued on next page}}\\
\endfoot
\bottomrule
\endlastfoot
MIMIC-CXR       & Pleural Effusion                    & Effusion                      & rename \\
MIMIC-CXR       & uncertain ($-1$)                    & \textit{missing}              & recode $\rightarrow$ NaN \\
\addlinespace
Nagoya          & Lung\_lesion                        & Lung Lesion                   & rename \\
Nagoya          & PleuralEffusion                     & Pleural Effusion              & rename \\
Nagoya          & LungOpacity                         & Lung Opacity                  & rename \\
Nagoya          & PleuralThickening                   & Pleural\_Thickening           & rename \\
Nagoya          & Cardiomediastinum                   & Enlarged Cardiomediastinum    & semantic \\
Nagoya          & Other\_abnormality                  & \textit{dropped}              & drop (free text) \\
\addlinespace
TAIX Ray        & PulmonaryCongestion                 & Edema                         & semantic (lossy) \\
TAIX Ray        & PleuralEffusion\_\{L,R\}            & Pleural Effusion              & rename + OR-merge \\
TAIX Ray        & PulmonaryOpacities\_\{L,R\}         & Lung Opacity                  & rename + OR-merge \\
TAIX Ray        & Atelectasis\_\{L,R\}               & Atelectasis                   & OR-merge \\
TAIX Ray        & grade $0$--$4$                      & \textit{binary}               & recode: \{0,1\}$\to$0, \{2,4\}$\to$1, else NaN \\
\addlinespace
Thorax XR       & nodule\_or\_mass                    & Nodule or mass                & rename \\
Thorax XR       & enlarged\_cardiomediastinum         & Enlarged Cardiomediastinum    & rename \\
Thorax XR       & lung\_lesion                        & Lung Lesion                   & rename \\
Thorax XR       & pleural\_thickening                 & Pleural\_Thickening           & rename \\
Thorax XR       & lung\_opacity                       & Lung Opacity                  & rename \\
Thorax XR       & not\_given / unknown                & \textit{missing}              & recode $\rightarrow$ NaN \\
\addlinespace
MIDRC           & COVID-19 pneumonia                  & \textit{excluded}             & drop (out of vocabulary) \\
\addlinespace
Jinling & 41 Chinese columns                & 36 canonical English          & translate \\
Jinling & positions 8, 9                    & Pleural Other                 & OR-merge \\
Jinling & positions 17, 18                  & Fracture                      & OR-merge \\
Jinling & positions 22, 25, 26              & \textit{dropped}              & drop (no English target) \\
\addlinespace
CheXpert$^{\dagger}$   & Pleural Effusion               & Effusion                      & rename \\
PadChest$^{\dagger}$   & Spanish finding synonyms       & e.g.\ Infiltration            & synonym map \\
VinDr-CXR$^{\dagger}$  & Pleural effusion               & Effusion                      & rename \\
IU-Xray$^{\dagger}$      & Infiltrate / Atelectases       & Infiltration / Atelectasis    & rename \\
\end{longtable}
\begin{flushleft}\footnotesize
$^{\dagger}$Inherited from the \texttt{torchxrayvision} library
(CheXpert, PadChest, VinDr-CXR, IU-Xray, ChestX-ray14 (Google)). Uncertain, unknown, or
unmappable labels are recoded to missing and excluded from scoring for every
dataset. Complete column-by-column provenance (\texttt{file:line}) is given in
\texttt{label\_harmonization\_raw.md}.
\end{flushleft}
\endgroup

\begingroup\small
\setlength{\tabcolsep}{5pt}
\renewcommand{\arraystretch}{1.15}
\begin{longtable}{@{}p{3.0cm}p{3.4cm}p{6.4cm}@{}}
\caption{Prediction-side alignment of model outputs to the standardized finding
vocabulary. Multiple model heads mapped to one finding are aggregated by their
maximum. Ground-truth labels are always read from the true label column; this
alignment applies only to model outputs.}
\label{tab:pred-harmonization}\\
\toprule
\textbf{Model(s)} & \textbf{Canonical finding} & \textbf{Model output heads / concepts} \\
\midrule
\endfirsthead
\multicolumn{3}{@{}l}{\emph{Table~\ref{tab:pred-harmonization} (continued)}}\\
\toprule
\textbf{Model(s)} & \textbf{Canonical finding} & \textbf{Model output heads / concepts} \\
\midrule
\endhead
\midrule
\multicolumn{3}{r@{}}{\emph{continued on next page}}\\
\endfoot
\bottomrule
\endlastfoot
\multicolumn{3}{@{}l}{\textit{Canonical-name alias table (metric-time; prediction side only)}}\\
\addlinespace
--              & Pleural Effusion            & Effusion \\
--              & Lung Lesion                 & Lesion \\
--              & Nodule or mass              & Nodule/Mass, Nodule, Mass \\
\addlinespace
\multicolumn{3}{@{}l}{\textit{Per-model concept maps (representative multi-head aggregations)}}\\
\addlinespace
Resnet, DenseNet & Nodule/Mass                & bulla, cavity, parenchymal\_lesion, apical\_changes \\
XRayDar         & Enlarged Cardiomediastinum  & mediastinum\_widened, cardiomegaly \\
XRayDar         & Pneumonia                   & bronchial\_changes, consolidation, ground\_glass\_opacification, interstitial\_shadowing, COVID-19 pneumonia \\
XRayDar         & Lung Opacity                & interstitial\_shadowing, ground\_glass\_opacification \\
XRayDar         & Lung Lesion / Nodule        & bulla, cavity, parenchymal\_lesion, apical\_changes \\
XRayDar         & Fracture                    & rib\_fracture, clavicle\_fracture \\
XRayDar         & Support Devices             & object \\
MAVL, MedKLIP   & Cardiomegaly                & cardiomegaly, enlarged cardiomediastinum \\
MAVL, MedKLIP   & Mass                        & nodule, nodule/mass, mass \\
MAVL, MedKLIP   & Pleural Other               & effusion \\
MAVL, MedKLIP   & ILD / Fibrosis / Granuloma  & disease / scar / nodule \\
CheXagent, CheXagent-AutoGluon & Pneumonia    & COVID-19 pneumonia \\
\end{longtable}
\begin{flushleft}\footnotesize
BioViL-T, CheXzero, and PsPG define no concept map and are aligned by
case-insensitive name matching plus the alias table above. Full provenance
(\texttt{file:line}) is given in \texttt{label\_harmonization\_raw.md}
(Section~3).
\end{flushleft}
\endgroup

%% file: tables/supS_dataset_overview_full.tex
\begin{table}[p]
\centering
\caption{Overview of the chest-radiograph datasets in the benchmark: access, provenance, image format, evaluated cohort size, and the per-task prevalence (\% positive in the evaluated cohort) of each imaging label. Complete version of table 1 of the Article: all 52 diagnostic tasks, including the 25 annotated by a single cohort.}
\label{tab:dataset-overview-full}
\scriptsize
\setlength{\tabcolsep}{3.5pt}
\renewcommand{\arraystretch}{1.0}
\begin{adjustbox}{max width=\textwidth,max totalheight=0.82\textheight,center}
\begin{tabular}{>{\raggedright\arraybackslash}p{3.0cm} !{\color{black!18}\vrule} c c c c c c c c c c c c}
\toprule
 & \multicolumn{7}{c}{\textcolor{pubcol}{\textsc{Public}}} & \multicolumn{5}{c}{\textcolor{pricol}{\textsc{Private}}} \\
\cmidrule(lr){2-8}\cmidrule(lr){9-13}
\textbf{Dataset} & \rotatebox{90}{\textbf{CheXpert}} & \rotatebox{90}{\textbf{MIDRC}} & \rotatebox{90}{\textbf{ChestX-ray14 (Google)}} & \rotatebox{90}{\textbf{IU-Xray}} & \rotatebox{90}{\textbf{PadChest}} & \rotatebox{90}{\textbf{VinDr-CXR}} & \rotatebox{90}{\textbf{MIMIC-CXR}} & \rotatebox{90}{\textbf{Nagoya}} & \rotatebox{90}{\textbf{Jinling}} & \rotatebox{90}{\textbf{Nanjing Multi-Hospital}} & \rotatebox{90}{\textbf{TAIX Ray}} & \rotatebox{90}{\textbf{Thorax XR}} \\
\midrule
\textbf{Country} & USA & USA & USA & USA & Spain & Vietnam & USA & Japan & China & China & Germany & Germany \\
\textbf{Image format} & JPEG & DICOM & PNG & PNG & PNG & DICOM & JPEG & DICOM & DICOM & DICOM & PNG & DICOM \\
\textbf{Images, total} & 224{,}316 & 6{,}676 & 112{,}120 & 7{,}470 & 160{,}868 & 18{,}000 & 377{,}110 & 1{,}201 & 161{,}175 & 2{,}529 & 215{,}385 & 5{,}269 \\
\textbf{Images, frontal} & 191{,}010 & 6{,}676 & 112{,}120 & \textemdash{} & \textemdash{} & 18{,}000 & 243{,}334 & 1{,}201 & 161{,}175 & 2{,}529 & 215{,}385 & 5{,}269 \\
\textbf{Official test, $n$} & 668 & \textemdash{} & 1{,}962 & \textemdash{} & \textemdash{} & 3{,}000 & 5{,}159 & \textemdash{} & \textemdash{} & \textemdash{} & \textemdash{} & \textemdash{} \\
\textbf{Evaluated, $n$} & \textbf{403} & \textbf{6{,}676} & \textbf{1{,}695} & \textbf{2{,}561} & \textbf{62{,}838} & \textbf{3{,}000} & \textbf{3{,}403} & \textbf{1{,}201} & \textbf{161{,}175} & \textbf{2{,}529} & \textbf{215{,}385} & \textbf{5{,}269} \\
\midrule
\multicolumn{13}{l}{\itshape Per-task prevalence (\% positive in evaluated cohort; blank = label not provided; ordered by number of datasets)} \\
\addlinespace[1pt]
Fracture & \cellcolor{prevcol!11}1.2 &  & \cellcolor{prevcol!19}3.5 & \cellcolor{prevcol!16}2.4 & \cellcolor{prevcol!16}2.7 &  & \cellcolor{prevcol!36}13 & \cellcolor{prevcol!50}25 & \cellcolor{prevcol!20}4.2 & \cellcolor{prevcol!30}8.9 & \cellcolor{prevcol!42}18 & \cellcolor{prevcol!62}82 \\
\rowcolor{rowband}Pleural Effusion & \cellcolor{prevcol!45}20 &  &  & \cellcolor{prevcol!22}4.9 & \cellcolor{prevcol!17}2.9 & \cellcolor{prevcol!19}3.7 & \cellcolor{prevcol!62}55 & \cellcolor{prevcol!60}36 & \cellcolor{prevcol!42}18 & \cellcolor{prevcol!33}11 & \cellcolor{prevcol!62}57 & \cellcolor{prevcol!62}41 \\
Pneumothorax & \cellcolor{prevcol!13}1.7 &  & \cellcolor{prevcol!21}4.2 & \cellcolor{prevcol!8}$<$1 & \cellcolor{prevcol!6}$<$1 & \cellcolor{prevcol!8}$<$1 & \cellcolor{prevcol!28}7.8 & \cellcolor{prevcol!44}19 & \cellcolor{prevcol!20}4.1 & \cellcolor{prevcol!26}7.0 &  & \cellcolor{prevcol!26}7.0 \\
\rowcolor{rowband}Atelectasis & \cellcolor{prevcol!55}31 &  &  & \cellcolor{prevcol!31}9.5 & \cellcolor{prevcol!20}4.1 & \cellcolor{prevcol!17}2.9 & \cellcolor{prevcol!62}56 & \cellcolor{prevcol!62}38 & \cellcolor{prevcol!8}$<$1 &  & \cellcolor{prevcol!62}38 & \cellcolor{prevcol!62}49 \\
Cardiomegaly & \cellcolor{prevcol!56}31 &  &  & \cellcolor{prevcol!26}6.8 & \cellcolor{prevcol!30}8.8 & \cellcolor{prevcol!32}10 & \cellcolor{prevcol!62}54 & \cellcolor{prevcol!37}13 & \cellcolor{prevcol!34}11 &  & \cellcolor{prevcol!62}85 & \cellcolor{prevcol!57}32 \\
\rowcolor{rowband}Lung Opacity & \cellcolor{prevcol!62}53 & \cellcolor{prevcol!62}80 & \cellcolor{prevcol!60}35 & \cellcolor{prevcol!33}11 &  & \cellcolor{prevcol!17}2.8 & \cellcolor{prevcol!62}63 & \cellcolor{prevcol!62}81 &  &  & \cellcolor{prevcol!62}40 & \cellcolor{prevcol!59}35 \\
Consolidation & \cellcolor{prevcol!25}6.2 &  &  &  & \cellcolor{prevcol!9}$<$1 & \cellcolor{prevcol!18}3.2 & \cellcolor{prevcol!48}23 & \cellcolor{prevcol!49}24 & \cellcolor{prevcol!9}$<$1 &  & \cellcolor{prevcol!62}78 & \cellcolor{prevcol!30}9.1 \\
\rowcolor{rowband}Edema & \cellcolor{prevcol!40}16 &  &  & \cellcolor{prevcol!14}2.1 & \cellcolor{prevcol!6}$<$1 &  & \cellcolor{prevcol!62}46 & \cellcolor{prevcol!6}$<$1 & \cellcolor{prevcol!6}$<$1 &  & \cellcolor{prevcol!62}55 & \cellcolor{prevcol!58}33 \\
Pneumonia & \cellcolor{prevcol!16}2.5 &  &  & \cellcolor{prevcol!14}1.9 & \cellcolor{prevcol!18}3.2 &  & \cellcolor{prevcol!55}30 & \cellcolor{prevcol!60}36 & \cellcolor{prevcol!53}28 & \cellcolor{prevcol!49}24 &  & \cellcolor{prevcol!15}2.3 \\
\rowcolor{rowband}Infiltration &  &  &  & \cellcolor{prevcol!17}2.8 & \cellcolor{prevcol!27}7.4 & \cellcolor{prevcol!14}1.9 &  & \cellcolor{prevcol!59}35 & \cellcolor{prevcol!6}$<$1 &  & \cellcolor{prevcol!37}14 & \cellcolor{prevcol!55}30 \\
Lung Lesion & \cellcolor{prevcol!11}1.2 &  &  & \cellcolor{prevcol!6}$<$1 &  & \cellcolor{prevcol!18}3.1 & \cellcolor{prevcol!41}17 & \cellcolor{prevcol!49}24 &  &  & \cellcolor{prevcol!62}90 & \cellcolor{prevcol!29}8.5 \\
\rowcolor{rowband}Enlarged cardiomediastinum & \cellcolor{prevcol!62}49 &  &  &  &  &  & \cellcolor{prevcol!42}18 & \cellcolor{prevcol!23}5.2 & \cellcolor{prevcol!6}$<$1 &  & \cellcolor{prevcol!62}48 & \cellcolor{prevcol!35}12 \\
Pleural Thickening &  &  &  & \cellcolor{prevcol!9}$<$1 & \cellcolor{prevcol!18}3.3 & \cellcolor{prevcol!24}5.6 &  & \cellcolor{prevcol!62}58 & \cellcolor{prevcol!29}8.6 &  &  & \cellcolor{prevcol!49}24 \\
\rowcolor{rowband}Emphysema &  &  &  & \cellcolor{prevcol!16}2.4 & \cellcolor{prevcol!9}$<$1 &  &  & \cellcolor{prevcol!48}23 & \cellcolor{prevcol!12}1.4 &  &  &  \\
Mass &  &  &  & \cellcolor{prevcol!6}$<$1 & \cellcolor{prevcol!9}$<$1 &  &  & \cellcolor{prevcol!34}12 & \cellcolor{prevcol!13}1.6 &  &  &  \\
\rowcolor{rowband}Nodule &  &  &  & \cellcolor{prevcol!14}1.9 & \cellcolor{prevcol!19}3.5 &  &  & \cellcolor{prevcol!53}28 & \cellcolor{prevcol!22}4.8 &  &  &  \\
Nodule or mass &  &  & \cellcolor{prevcol!37}14 &  &  & \cellcolor{prevcol!24}5.9 &  &  &  &  & \cellcolor{prevcol!62}61 & \cellcolor{prevcol!29}8.2 \\
\rowcolor{rowband}Support Devices & \cellcolor{prevcol!62}51 &  &  &  & \cellcolor{prevcol!14}1.8 &  & \cellcolor{prevcol!62}63 &  & \cellcolor{prevcol!15}2.3 &  &  &  \\
Fibrosis &  &  &  & \cellcolor{prevcol!7}$<$1 & \cellcolor{prevcol!7}$<$1 &  &  &  & \cellcolor{prevcol!6}$<$1 &  &  &  \\
\rowcolor{rowband}Hernia &  &  &  & \cellcolor{prevcol!12}1.5 & \cellcolor{prevcol!13}1.6 &  &  &  & \cellcolor{prevcol!6}$<$1 &  &  &  \\
Pleural Other & \cellcolor{prevcol!9}$<$1 &  &  &  &  &  & \cellcolor{prevcol!31}9.6 &  & \cellcolor{prevcol!8}$<$1 &  &  &  \\
\rowcolor{rowband}Aortic Atherosclerosis &  &  &  &  & \cellcolor{prevcol!14}2.1 &  &  &  & \cellcolor{prevcol!62}44 &  &  &  \\
Aortic Enlargement &  &  &  &  &  & \cellcolor{prevcol!27}7.3 &  &  & \cellcolor{prevcol!12}1.4 &  &  &  \\
\rowcolor{rowband}Calcification &  &  &  &  &  & \cellcolor{prevcol!25}6.5 &  &  & \cellcolor{prevcol!17}2.8 &  &  &  \\
Granuloma &  &  &  & \cellcolor{prevcol!32}10 & \cellcolor{prevcol!16}2.5 &  &  &  &  &  &  &  \\
\rowcolor{rowband}ILD &  &  &  &  &  & \cellcolor{prevcol!27}7.4 &  &  & \cellcolor{prevcol!21}4.4 &  &  &  \\
Scoliosis &  &  &  &  & \cellcolor{prevcol!25}6.3 &  &  &  & \cellcolor{prevcol!12}1.3 &  &  &  \\
\rowcolor{rowband}Air Trapping &  &  &  &  & \cellcolor{prevcol!19}3.6 &  &  &  &  &  &  &  \\
Aortic Elongation &  &  &  &  & \cellcolor{prevcol!29}8.2 &  &  &  &  &  &  &  \\
\rowcolor{rowband}Bone Destruction &  &  &  &  &  &  &  &  & \cellcolor{prevcol!6}$<$1 &  &  &  \\
Bronchiectasis &  &  &  &  & \cellcolor{prevcol!12}1.4 &  &  &  &  &  &  &  \\
\rowcolor{rowband}Calcified Granuloma &  &  &  & \cellcolor{prevcol!24}5.7 &  &  &  &  &  &  &  &  \\
Chronic Bronchitis &  &  &  &  &  &  &  &  & \cellcolor{prevcol!10}$<$1 &  &  &  \\
\rowcolor{rowband}Costophrenic Angle Blunting &  &  &  &  & \cellcolor{prevcol!18}3.4 &  &  &  &  &  &  &  \\
COVID-19 pneumonia &  & \cellcolor{prevcol!51}26 &  &  &  &  &  &  &  &  &  &  \\
\rowcolor{rowband}Degeneration &  &  &  &  &  &  &  &  & \cellcolor{prevcol!8}$<$1 &  &  &  \\
Flattened Diaphragm &  &  &  &  & \cellcolor{prevcol!7}$<$1 &  &  &  &  &  &  &  \\
\rowcolor{rowband}Ground Glass Opacity &  &  &  &  &  &  &  &  & \cellcolor{prevcol!20}3.9 &  &  &  \\
Healed Tuberculosis &  &  &  &  &  &  &  &  & \cellcolor{prevcol!20}3.9 &  &  &  \\
\rowcolor{rowband}Hemidiaphragm Elevation &  &  &  &  & \cellcolor{prevcol!12}1.4 &  &  &  &  &  &  &  \\
Hilar Enlargement &  &  &  &  & \cellcolor{prevcol!21}4.6 &  &  &  &  &  &  &  \\
\rowcolor{rowband}Incomplete lung expansion &  &  &  &  &  &  &  &  & \cellcolor{prevcol!6}$<$1 &  &  &  \\
Non-pulmonary Mass &  &  &  &  &  &  &  &  & \cellcolor{prevcol!6}$<$1 &  &  &  \\
\rowcolor{rowband}Normal &  &  &  &  &  &  &  &  & \cellcolor{prevcol!38}15 &  &  &  \\
Parenchymal Band &  &  &  &  &  &  &  &  & \cellcolor{prevcol!21}4.4 &  &  &  \\
\rowcolor{rowband}PICC (central catheter) &  &  &  &  &  &  &  &  & \cellcolor{prevcol!29}8.2 &  &  &  \\
Pneumoperitoneum &  &  &  &  &  &  &  &  & \cellcolor{prevcol!7}$<$1 &  &  &  \\
\rowcolor{rowband}Pulmonary Arterial Hypertension &  &  &  &  &  &  &  &  & \cellcolor{prevcol!6}$<$1 &  &  &  \\
Pulmonary Fibrosis &  &  &  &  &  & \cellcolor{prevcol!27}7.2 &  &  &  &  &  &  \\
\rowcolor{rowband}Subcutaneous Emphysema &  &  &  &  &  &  &  &  & \cellcolor{prevcol!16}2.6 &  &  &  \\
Tube &  &  &  &  & \cellcolor{prevcol!10}$<$1 &  &  &  &  &  &  &  \\
\rowcolor{rowband}Tuberculosis &  &  &  &  & \cellcolor{prevcol!8}$<$1 &  &  &  &  &  &  &  \\
\bottomrule
\end{tabular}%
\end{adjustbox}

\smallskip
\begin{minipage}{\textwidth}\scriptsize
\emph{Images, total/frontal} are the full published dataset sizes; \emph{Official test, $n$} is the dataset's official held-out test split (em-dash = no official split); \emph{Evaluated, $n$} is the cohort scored in this study (distinct evaluated images; frontal AP/PA where view information was recorded). For private cohorts total = frontal = evaluated. \emph{Per-task prevalence} is the percentage of the evaluated cohort positive for that label ($n_{\mathrm{pos}}/n_{\mathrm{valid}}$); cells are shaded by prevalence (darker = more prevalent); ``$<$1'' denotes a non-zero prevalence below 1\%. For the pooled Nanjing Multi-Hospital column prevalence is over the combined subsets. Synonymous label names are merged to a single row (e.g.\ Effusion/Pleural~Effusion; ILD variants; Supportive~Device/Support~Devices). Dataset access (public versus private), country, and image format are given in the column grouping and header rows. MIDRC-RICORD-1C's pneumonia label is COVID-19 pneumonia; ChestX-ray14 (Google) carries only the four Google-adjudicated findings (no Pneumonia). TAIX~Ray and Thorax~XR are two distinct German cohorts; the three Nanjing Multi-Hospital subsets are pooled. TAIX~Ray was publicly released in February 2026 under a CC~BY~4.0 licence, which post-dates the development of all 10 evaluated models; it is grouped with the private cohorts because no evaluated model can have been trained on it, not because it is unavailable. Thorax~XR has no public release.
\end{minipage}
\end{table}

%% file: sections/S2_models.tex
\section{Summary of evaluated models}\label{sec:supp-models}

We evaluated ten chest X-ray models spanning the four learning paradigms of the main text (table 2 of the Article): supervised classifiers, vision-language foundation models, an instruction-tuned QA model, and a repurposed foundation-model encoder. Within the vision-language group the models differ further by training strategy---contrastive, knowledge-enhanced alignment, and prompt learning---as reflected in the objective column of table 2. This section gives a narrative account of that table. Two supervised convolutional baselines, DenseNet-121 and ResNet-50, were taken from the TorchXRayVision library and trained on the union of public chest X-ray datasets with harmonized labels; X-Raydar is a multi-scale convolutional ensemble trained on a large multi-hospital cohort with labels extracted from reports. Two contrastive vision-language models, CheXzero and BioViL-T, learn a shared image-text embedding from paired radiographs and reports and classify findings zero-shot by comparing an image against natural-language prompts. Two knowledge-enhanced models, MedKLIP and MAVL, replace free-text prompts with structured supervision: MedKLIP aligns images to report-derived entity triplets against a medical knowledge base, and MAVL decomposes each disease into interpretable visual aspects derived from a large language model. PsPG adapts a pretrained vision-language model by prompt learning, generating class-specific pseudo-prompts for multi-label recognition. Finally, two generative or repurposed foundation-model approaches were evaluated: CheXagent, an instruction-tuned vision-language model that answers natural-language questions about a radiograph and whose free-text output is parsed into a label, and CheXagent-AutoGluon, which repurposes the frozen CheXagent image encoder as a feature extractor for a supervised tabular ensemble (\cref{sec:supp-autogluon}).

These paradigms differ both in the data they are trained on and in how a prediction is obtained---label logits, text-prompt similarity, or parsed text generation. Because CheXagent emits categorical Yes/No answers rather than calibrated probabilities, it is excluded from the probability-based operating-point analyses but retained as an evaluated model. The architectures, training data, objectives, inference interfaces, and source references for all ten models are summarized in table 2 of the Article.

%% file: sections/S3_autogluon.tex
\section{The CheXagent-AutoGluon classifier}\label{sec:supp-autogluon}

\paragraph{Feature extraction.}
The CheXagent-AutoGluon (``CheXagent-AG'') model repurposes the CheXagent-8b foundation model (\texttt{StanfordAIMI/CheXagent-8b}) as a frozen image feature extractor. For each radiograph, only the vision encoder submodule of CheXagent-8b was used; the language decoder was not invoked. Images were preprocessed with the model's native processor and resized to $448 \times 448$, and the encoder's pooled representation was taken as a single 1{,}408-dimensional embedding per image. All feature extraction was performed in half precision (float16) with the encoder weights held fixed (no fine-tuning of the CheXagent backbone).

\paragraph{Training data.}
Embeddings were extracted for the CheXpert training set (225{,}008 radiographs, comprising both frontal and lateral views). Fourteen findings were considered: Enlarged Cardiomediastinum, Cardiomegaly, Lung Opacity, Lung Lesion, Edema, Consolidation, Pneumonia, Atelectasis, Pneumothorax, Pleural Effusion, Pleural Other, Fracture, Support Devices, and an aggregate ``Any Abnormalities'' label (defined as the complement of the CheXpert ``No Finding'' label, i.e.\ positive when any finding is present). Each finding was binarized independently by mapping the positive CheXpert label to 1 and all other values (including uncertain, $-1$, and blank labels) to 0. Positive prevalence in the training set ranged from approximately 1.8\% (Pleural Other) and 2.2\% (Pneumonia) to 50.6\% (Support Devices) and 88.7\% (Any Abnormalities), reflecting substantial class imbalance across findings.

\paragraph{Model training.}
For each finding, a separate binary classifier was trained on the 1{,}408-dimensional embeddings using AutoGluon's \texttt{TabularPredictor} (autogluon.tabular v1.2). Training used the \texttt{high\_quality} preset together with \texttt{optimize\_for\_deployment}, which fits a stacked ensemble of tabular learners (including tree-based models such as ExtraTrees and neural-network models with hyperparameter search) with 8-fold bagging and out-of-fold cross-validation, then refits the selected models on the full training data and prunes the ensemble to a compact set of deployed members combined by a weighted-ensemble meta-learner. The default binary evaluation metric (classification accuracy) was used for model selection; no separate held-out tuning set was supplied to the fitting routine, so model validation relied on AutoGluon's internal cross-validation folds. The procedure yielded one predictor per finding.

\paragraph{Inference.}
At evaluation time, each test radiograph was passed through the frozen CheXagent-8b vision encoder to obtain its 1{,}408-dimensional embedding, which was assembled into a tabular feature vector. For each finding, the corresponding AutoGluon predictor produced a positive-class probability, which was used directly as the continuous score for that finding (no additional calibration or rescaling was applied). Scores were generated for the 13 deployed findings for which a predictor was included at inference time.

%% file: sections/S4_performance.tex
\section{Model performance profiles across diagnostic categories}\label{sec:supp-performance}

\begin{figure}[p]
    \centering
    \includegraphics[width=\linewidth,height=0.9\textheight,keepaspectratio]{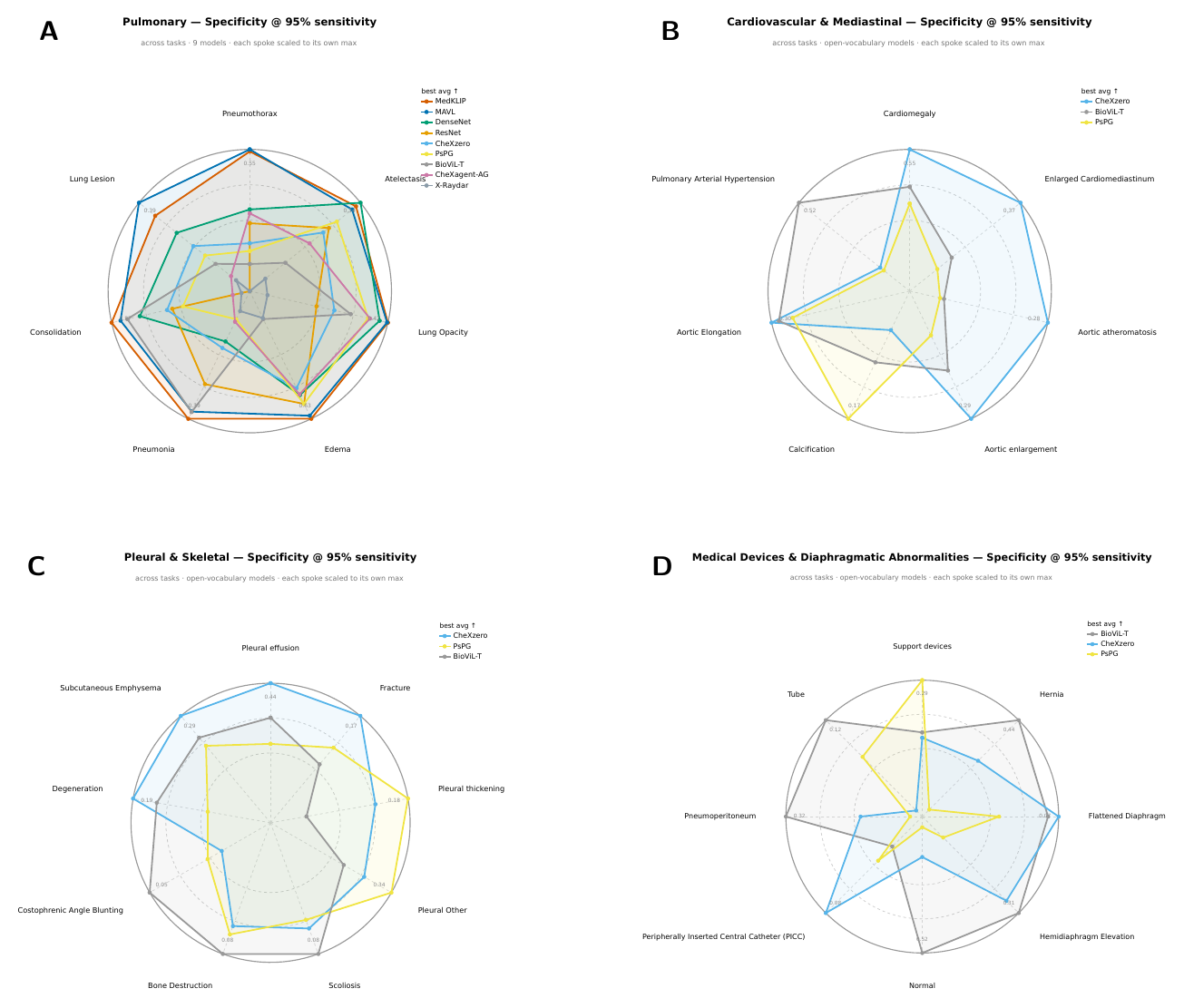}
    \caption{\textbf{Per-category performance profiles across diagnostic tasks.} Radar charts of specificity at 95\% sensitivity across the individual tasks within each anatomy category: \textbf{A}, pulmonary; \textbf{B}, cardiovascular and mediastinal; \textbf{C}, pleural and skeletal; \textbf{D}, medical devices and diaphragmatic abnormalities. The pulmonary category (\textbf{A}) contains enough tasks for all nine models to share more than five gap-free spokes, so the full model fleet is shown; the remaining categories (\textbf{B}--\textbf{D}) have too few tasks for the fixed-vocabulary models to complete a shared spoke set, so only the three open-vocabulary models (CheXzero, BioViL-T, PsPG) are drawn, and full-model comparison for these categories is given in Figure~4 and Figure~5B. Each spoke is a canonical finding; categories with more than twelve tasks show the twelve best-covered spokes. Within each spoke, values are scaled to the maximum across the plotted models, so the radars show relative profiles rather than absolute specificity; the annotated value gives the top score on that spoke.}
    \label{fig:supp-radar}
\end{figure}

The main-text analyses summarize task-level performance by diagnostic category as aggregate bars (Figure~5B) and as a model-by-task heatmap (Figure~4). To complement these with a per-model view of the \emph{shape} of performance within a category---which findings a model is relatively strong or weak on, rather than its category average---we plot radar profiles of specificity at 95\% sensitivity across the individual tasks that make up each anatomy category (\cref{fig:supp-radar}).

A radar requires every plotted model to score every spoke, or the polygon breaks. The open-vocabulary models (CheXzero, BioViL-T, and PsPG) emit a score for an arbitrary finding name and therefore cover every task in a category; the remaining models predict only their fixed label vocabulary and complete only a fraction of a category's tasks. Whether a full-fleet radar is feasible therefore depends on how many tasks a category contains. The pulmonary category is task-rich enough (28 canonical findings) that all nine models share more than five gap-free spokes, so its radar (panel A) shows the complete model fleet. The three remaining categories contain too few tasks for the fixed-vocabulary models to complete a shared spoke set, so their radars (panels B--D) retain the three open-vocabulary models; full-model comparison for these categories is instead reported in the main-text bar charts (Figure~5B) and heatmap (Figure~4). Tasks are grouped into four anatomy categories (pulmonary; cardiovascular and mediastinal; pleural and skeletal; medical devices and diaphragmatic abnormalities), and categories with more than twelve tasks show the twelve best-covered spokes. Within each spoke, values are scaled to the maximum across the plotted models, so the radars display the \emph{relative} profile of the models on each finding rather than absolute specificity; the annotated value on each spoke gives the corresponding top score. Consistent with the overall results, no single model dominated across findings: relative strengths shifted from task to task within every category, reinforcing the task-level heterogeneity described in the main text.

%% file: sections/S4b_spatial_extent.tex
\section{Definition of the spatial-extent categories}\label{sec:supp-extent}

The spatial-extent stratification (figure 5C of the Article) groups each diagnostic task by how much of the radiograph the finding typically occupies and how spatially conspicuous it is, on the clinical intuition that small, high-spatial-frequency abnormalities (for example a fracture or a lung nodule) are easy to miss whereas large-area changes (for example widespread opacity or cardiomegaly) dominate the image and are more readily detected. The assignment is a deliberate, \emph{a priori} clinical label applied to the canonical finding vocabulary; it is deterministic and fixed before any metric is computed, so extent can be read against performance without circularity. It is not derived from any model output, prevalence, or accuracy. The three axes used in this work---anatomy (\cref{sec:supp-performance}), spatial extent, and prevalence (\cref{sec:supp-prevalence})---are treated as orthogonal, so a task's extent is independent of its anatomical home and of how common it is (for example, aortic atherosclerosis is focal in extent but common in prevalence, whereas pneumonia is regional in extent and also common).

Each canonical task was assigned to exactly one of three levels:

\begin{itemize}
  \item \textbf{Focal} --- small, high-spatial-frequency findings that occupy a small fraction of the field and are easy to miss (17 tasks): Aortic atherosclerosis, Bone Destruction, Calcification, Calcified Granuloma, Costophrenic Angle Blunting, Fracture, Granuloma, Hilar Enlargement, Mass, Nodule, Nodule/Mass, Non-pulmonary Mass, Parenchymal Band, PICC, Pneumoperitoneum, Subcutaneous Emphysema, and Tube.
  \item \textbf{Regional} --- findings occupying a definable segment, lobe, or anatomical territory (25 tasks): Air Trapping, Aortic Elongation, Aortic Enlargement, Atelectasis, Bronchiectasis, Chronic Bronchitis, Consolidation, COVID-19 pneumonia, Degeneration, Enlarged Cardiomediastinum, Flattened Diaphragm, Ground Glass Opacity, Healed Tuberculosis, Hemidiaphragm Elevation, Hernia, Incomplete Lung Expansion, Infiltration, Lesion, Lung Lesion, Pleural Other, Pleural Thickening, Pneumonia, Pneumothorax, Scoliosis, and Tuberculosis.
  \item \textbf{Diffuse} --- large-area findings that dominate the field and are conspicuous (10 tasks): Cardiomegaly, Edema, Emphysema, Fibrosis, ILD, Lung Opacity, Normal, Pleural Effusion, Pulmonary Arterial Hypertension, and Support Devices.
\end{itemize}

Consistent with the main-text result that specificity at 95\% sensitivity was lowest for focal, intermediate for regional, and highest for diffuse findings (figure 5C of the Article), this ordering reflects detectability rather than any dataset-specific artefact: the categories are defined purely from the anatomy and typical appearance of each finding.

%% file: sections/S5_prevalence.tex
\section{Definition of the common versus rare prevalence threshold}\label{sec:supp-prevalence}

Because our primary endpoints are operating-point statistics evaluated at a fixed high-sensitivity threshold (specificity and positive predictive value at 95\% sensitivity), their stability depends directly on the number of positive examinations available for each diagnostic task. For very low-prevalence findings the positive count in a cohort can fall to a few dozen examinations or fewer (as low as 3--21 positives in our data; \cref{tab:task-prevalence}), so the resulting per-task estimates are dominated by sampling noise and, in the case of PPV, are structurally bounded near the prevalence itself and degenerate toward zero as prevalence falls. To characterize this long tail without over-interpreting unstable per-task numbers, we defined a prevalence stratum \emph{a priori} and split diagnostic tasks into common and rare at a 1\% positive-prevalence threshold.

Task prevalence was computed on the harmonized evaluation cohorts as the fraction of positive examinations, taken once per (task, cohort) and then averaged with equal weight across the cohorts that evaluate the task, so that the classification is not dominated by any single large dataset; the three Nanjing Multi-Hospital subsets were pooled into one cohort beforehand. The 1\% value is a round, conventional boundary for genuinely uncommon findings and isolates the long-tail tasks from the bulk of the distribution: a task was labelled rare if its cohort-averaged prevalence fell below 1\% and common otherwise. This threshold gates only the prevalence-stratified panel of the diagnostic-heterogeneity analysis (Figure~5D); all other analyses use prevalence, where relevant, as a continuous quantity.

Across 166 evaluable dataset--task units, 32 (19\%) fell below the 1\% threshold (rare) and 134 were common. At the task level (cohort-averaged prevalence, the quantity used in Figure~5D), 10 of 52 canonical tasks were rare and 42 were common. Every rare task was rare in only the single cohort that evaluated it, and the rare dataset--task cells were concentrated in the large low-prevalence cohorts, chiefly Jinling and PadChest, where many fine-grained findings occur in well under 1\% of studies. Several tasks that are rare in a specific cohort are common on average across datasets (for example Pneumothorax, Edema, Fibrosis, and Consolidation), which is why the classification for the figure is made on the cohort-averaged prevalence rather than per cell. The full task-level classification is given in \cref{tab:task-prevalence}.

\input{tables/supS_task_prevalence}

%% file: tables/supS_task_prevalence.tex
\newcommand{\rareflag}{\textbf{\textcolor{BrickRed}{Rare}}}

\begingroup\small\setlength{\tabcolsep}{4pt}
\begin{longtable}{@{}l r r r r l@{}}
\caption{Diagnostic tasks classified as common versus rare at a 1\% positive-prevalence
threshold. Prevalence is the fraction of positive examinations, computed once per
(task, cohort) and averaged with equal weight across the cohorts evaluating the task
(Nanjing Multi-Hospital subsets pooled). \emph{\# Datasets} is the number of cohorts
evaluating the task; \emph{Common}/\emph{Rare cells} count how many of those cohorts
are $\ge$1\% vs $<$1\%. \emph{Class} is the task-level label used in the
prevalence-stratified panel (Figure~5D): rare if the cohort-averaged prevalence is
below 1\%.}\label{tab:task-prevalence}\\
\toprule
Task & \# Datasets & Common & Rare & Prev.\ (\%) & Class \\
\midrule
\endfirsthead
\multicolumn{6}{@{}l}{\small\itshape Table~\ref{tab:task-prevalence} (continued)}\\
\toprule
Task & \# Datasets & Common & Rare & Prev.\ (\%) & Class \\
\midrule
\endhead
\midrule \multicolumn{6}{r@{}}{\small continued on next page}\\
\endfoot
\bottomrule
\endlastfoot
Non-pulmonary Mass & 1 & 0 & 1 & 0.04 & \rareflag \\
Incomplete Expansion of the Lung & 1 & 0 & 1 & 0.12 & \rareflag \\
Pulmonary Arterial Hypertension & 1 & 0 & 1 & 0.17 & \rareflag \\
Bone Destruction & 1 & 0 & 1 & 0.34 & \rareflag \\
Pneumoperitoneum & 1 & 0 & 1 & 0.43 & \rareflag \\
Flattened Diaphragm & 1 & 0 & 1 & 0.50 & \rareflag \\
Degeneration & 1 & 0 & 1 & 0.57 & \rareflag \\
Tuberculosis & 1 & 0 & 1 & 0.64 & \rareflag \\
Tube & 1 & 0 & 1 & 0.91 & \rareflag \\
Chronic Bronchitis & 1 & 0 & 1 & 0.94 & \rareflag \\
Hernia & 3 & 2 & 1 & 1.04 & Common \\
Bronchiectasis & 1 & 1 & 0 & 1.38 & Common \\
Hemidiaphragm Elevation & 1 & 1 & 0 & 1.45 & Common \\
Fibrosis & 4 & 1 & 3 & 2.09 & Common \\
Subcutaneous Emphysema & 1 & 1 & 0 & 2.62 & Common \\
Lesion & 1 & 1 & 0 & 3.13 & Common \\
Costophrenic Angle Blunting & 1 & 1 & 0 & 3.41 & Common \\
Mass & 4 & 2 & 2 & 3.56 & Common \\
Air Trapping & 1 & 1 & 0 & 3.65 & Common \\
Pleural Other & 3 & 1 & 2 & 3.67 & Common \\
Scoliosis & 2 & 2 & 0 & 3.82 & Common \\
Healed tuberculosis & 1 & 1 & 0 & 3.94 & Common \\
Ground glass opacity & 1 & 1 & 0 & 3.94 & Common \\
Parenchymal Band & 1 & 1 & 0 & 4.35 & Common \\
Aortic enlargement & 2 & 2 & 0 & 4.36 & Common \\
Hilar Enlargement & 1 & 1 & 0 & 4.59 & Common \\
Calcification & 2 & 2 & 0 & 4.63 & Common \\
Pneumothorax & 10 & 7 & 3 & 5.26 & Common \\
Calcified Granuloma & 1 & 1 & 0 & 5.74 & Common \\
ILD & 2 & 2 & 0 & 5.91 & Common \\
Granuloma & 2 & 2 & 0 & 6.43 & Common \\
Emphysema & 4 & 3 & 1 & 6.89 & Common \\
Aortic Elongation & 1 & 1 & 0 & 8.19 & Common \\
Peripherally Inserted Central Catheter (PICC) & 1 & 1 & 0 & 8.20 & Common \\
Nodule/Mass & 3 & 3 & 0 & 9.25 & Common \\
Nodule & 4 & 4 & 0 & 9.52 & Common \\
Consolidation & 7 & 5 & 2 & 9.62 & Common \\
Lung Lesion & 5 & 4 & 1 & 10.08 & Common \\
Infiltration & 6 & 5 & 1 & 12.88 & Common \\
Pneumonia & 8 & 8 & 0 & 14.52 & Common \\
Normal & 1 & 1 & 0 & 14.78 & Common \\
Fracture & 9 & 9 & 0 & 15.90 & Common \\
Pleural thickening & 6 & 5 & 1 & 16.68 & Common \\
Enlarged Cardiomediastinum & 5 & 4 & 1 & 16.94 & Common \\
Edema & 8 & 5 & 3 & 18.50 & Common \\
Cardiomegaly & 8 & 8 & 0 & 21.08 & Common \\
Aortic atheromatosis & 2 & 2 & 0 & 23.07 & Common \\
Pleural effusion & 10 & 10 & 0 & 24.82 & Common \\
COVID-19 pneumonia & 1 & 1 & 0 & 26.18 & Common \\
Support devices & 4 & 4 & 0 & 29.67 & Common \\
Atelectasis & 9 & 8 & 1 & 29.96 & Common \\
Lung Opacity & 9 & 9 & 0 & 44.79 & Common \\
\end{longtable}

\bigskip

\begin{longtable}{@{}l l r r r@{}}
\caption{The 32 rare dataset--task cells (cohort prevalence $<$1\%). These are
the long-tail cells that motivate the threshold; they are concentrated in the large
low-prevalence cohorts (chiefly Jinling and PadChest).}\label{tab:rare-cells}\\
\toprule
Task & Dataset & $n$ & Positives & Prevalence (\%) \\
\midrule
\endfirsthead
\toprule
Task & Dataset & $n$ & Positives & Prevalence (\%) \\
\midrule
\endhead
\bottomrule
\endlastfoot
Infiltration & Jinling & 161{,}056 & 21 & 0.013 \\
Hernia & Jinling & 161{,}056 & 55 & 0.034 \\
Non-pulmonary Mass & Jinling & 161{,}056 & 57 & 0.035 \\
Fibrosis & Jinling & 161{,}056 & 148 & 0.092 \\
Incomplete Expansion of the Lung & Jinling & 161{,}056 & 186 & 0.115 \\
Lung Lesion & IU-Xray & 2{,}561 & 3 & 0.117 \\
Edema & Jinling & 161{,}056 & 222 & 0.138 \\
Pneumothorax & PadChest & 62{,}838 & 104 & 0.166 \\
Pulmonary Arterial Hypertension & Jinling & 161{,}056 & 276 & 0.171 \\
Enlarged Cardiomediastinum & Jinling & 161{,}056 & 310 & 0.192 \\
Mass & IU-Xray & 2{,}561 & 6 & 0.234 \\
Edema & PadChest & 62{,}838 & 191 & 0.304 \\
Bone Destruction & Jinling & 161{,}056 & 546 & 0.339 \\
Edema & Nagoya & 1{,}201 & 5 & 0.416 \\
Pneumoperitoneum & Jinling & 161{,}056 & 689 & 0.428 \\
Flattened Diaphragm & PadChest & 62{,}838 & 312 & 0.497 \\
Fibrosis & IU-Xray & 2{,}561 & 13 & 0.508 \\
Fibrosis & PadChest & 62{,}838 & 343 & 0.546 \\
Degeneration & Jinling & 161{,}056 & 917 & 0.569 \\
Pneumothorax & VinDr-CXR & 3{,}000 & 18 & 0.600 \\
Atelectasis & Jinling & 161{,}056 & 972 & 0.604 \\
Pneumothorax & IU-Xray & 2{,}561 & 16 & 0.625 \\
Tuberculosis & PadChest & 62{,}838 & 401 & 0.638 \\
Pleural Other & Jinling & 161{,}056 & 1108 & 0.688 \\
Pleural Other & CheXpert & 403 & 3 & 0.744 \\
Pleural thickening & IU-Xray & 2{,}561 & 20 & 0.781 \\
Consolidation & Jinling & 161{,}056 & 1283 & 0.797 \\
Mass & PadChest & 62{,}838 & 508 & 0.808 \\
Consolidation & PadChest & 62{,}838 & 510 & 0.812 \\
Emphysema & PadChest & 62{,}838 & 557 & 0.886 \\
Tube & PadChest & 62{,}838 & 570 & 0.907 \\
Chronic Bronchitis & Jinling & 161{,}056 & 1514 & 0.940 \\
\end{longtable}
\endgroup